# The Search for Directed Intelligence

Philip Lubin

lubin@deepspace.ucsb.edu

Physics Dept. UCSB
Broida Hall, Room 2015C
Santa Barbara, CA 93106-9530

## ABSTRACT

We propose a search for sources of directed energy systems such as those now becoming technologically feasible on Earth. Recent advances in our own abilities allow us to foresee our own capability that will radically change our ability to broadcast our presence. We show that systems of this type have the ability to be detected at vast distances and indeed can be detected across the entire horizon. This profoundly changes the possibilities for searches for extra-terrestrial technology advanced civilizations. We show that even modest searches can be extremely effective at detecting or limiting many civilization classes. We propose a search strategy, using small Earth based telescopes, that will observe more than $10^{12}$ stellar and planetary systems with possible extensions to more than $10^{20}$ systems allowing us to test the hypothesis that other similarly or more advanced civilization with this same capability, and are broadcasting, exist. We show that such searches have unity probability of detecting even a single comparably advanced civilization anywhere in our galaxy within a relatively short search time (few years) IF that civilization adopts a simple beacon strategy we call "intelligent targeting", IF that civilization is beaconing at a wavelength we can detect and IF that civilization left the beacon on long enough for the light to reach us now. In this blind beacon and blind search strategy the civilization does not need to know where we are nor do we need to know where they are. This same basic strategy can be extended to extragalactic distances.

## 1. INTRODUCTION

One of humanities most profound questions is "are we alone". This continues to literally obsess much of humanity from the extremely diverse backgrounds and interests from scientific, philosophical and theological. Proof of the existence of other forms of life would greatly influence all of humanity. The great difficulty in finding life is that our physical exploration (planets physically explored) is woefully inadequate with a fractional search currently of order $10^{-20}$ since the number of planets, based on the recent Kepler data and the estimated number of stars, in our universe is estimated to be of order $10^{20-24}$ and we have visited of order unity planets. For the foreseeable future we lack the ability to physically search much beyond this. With remote sensing, as has been the domain of traditional SETI programs, we can greatly expand this search fraction assuming that there are other civilizations with comparable or greater technological evolution to our own AND that such civilizations are actively seeking detection in parts of the electromagnetic spectrum we can search in. All such remote sensing searches require us to make assumptions that may have no basis in reality. Hence the great difficulty in converting searches to statements on the existence of life beyond our own. But it is all we have to go on and hence it should be pursued consistent with reasonable levels of effort. A detection would forever change humanity while an upper limit based on our assumptions has only a modest effect. This is truly a "high risk, high payoff" area of inquiry and always has been. As always we are "now" centric and "anthropomorphic" centric in that we expect all other advanced civilizations to be like minded in their desire to answer the same profound question AND to go about searching in a similar manner. However, if all civilizations "listened" but did not "speak" there would be a profound universal silence. Hopefully, other advanced civilizations do not share our relative silence. A serious and important question is to envision our time evolution of detection by other civilizations. Our ability to seriously ponder the issue of remote sensing of life has only become possible in the last 100 years. This represents about 1% of civilized human existence , less than 0.1% of total human existence, less than $10^{-7}$ of life on Earth and less than $10^{-8}$ since the first stars and galaxies formed. While predictions are fraught with uncertainty, especially those concerning the future, it is somewhat easier to look into the recent past at our technological progress in relevant areas.

## 2. TECHNOLOGICAL DEVELOPMENT

One of the enabling technologies that is relevant is the extremely dramatic progress in solid state lasers and in particular to laser amplifiers that can be arrayed into larger elements. The latter point is the analog of phased array radar that is becoming more common. An analogous revolution is taking place in visible and near IR coherent systems allowing for free space beam combining with no upper limit to power. This is very much analogous to the revolution in computing that has been brought about by parallel processing where large arrays of modest processors are now ubiquitous for super computing with no upper limit to computation. There is a very close analogy both technologically and in system design to the use of large arrays of modest phased arrays (parallel processing) lasers to form an extremely large directed energy system. Indeed the typical doubling time for performance in the semiconductor computational domain per computational element (CPU) is approximately 1.5-2 years over nearly 5 decades of time. We plot the power from CW fiber lasers as an analog to the CPU, and see the doubling time over the last 25 years has been approximately 1.7 years or 20 months. This is remarkably similar to "Moore's Law" and has not hit a plateau yet. CPU speed hit a plateau for Si devices nearly a decade ago and the path forward has been to increase the number of processors – ie to go toward parallel computer. You are likely reading this on such a CPU. Our current technology (early 2015) is above 1 Kw in a single mode fiber per amplifier with the analog of multi core CPU's being multi spectral injection with many fiber amplifiers per single mode fiber which now exceeds 30 Kw per fiber. It is estimated that this can be pushed to beyond 100 Kw per single mode fiber in the near future. We assume that other civilization possess the basic technology of arrayed (parallel) directed energy systems below but we only assume 1 Kw per fiber that we have already achieved. The efficiency of laser amplifiers is nearly 50% and thus only modest efficiency improvement is possible since we are already within a factor of two of unity. The power density is currently at about 5kg/kw and will drop to about 1 kg/kw in the next few years. All of this is a remarkable statement about our current technological capability in directed energy systems. As we will see we now possess the capability to deploy this technology in a way that enables us to direct energy for revolutionary purposes one of which is to be "seen" across the entire universe. This is truly a remarkable statement. The question that is relevant here is "if there are other advanced civilization do they have similar capabilities" and if so are they directing it to us? We have never been in a technological state where we could make such a statement and hence it is logical to explore its ramifications in many areas, SETI being one of them.

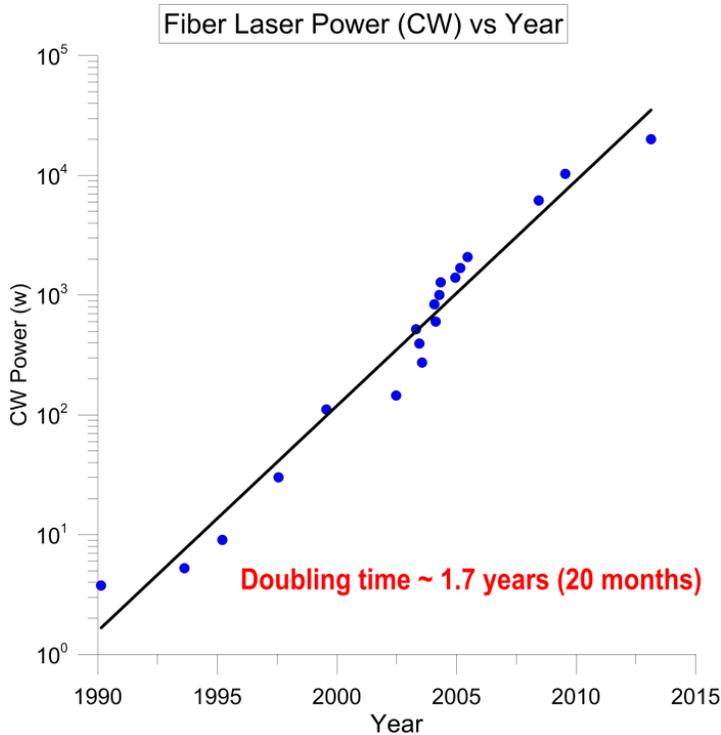

**Figure 1** – Fiber laser CW output power vs year over the past 25 years based on data in the literature.

## 3. CIVILIZATION CLASSES AND SIGNAL LEVEL

All SETI programs require assumptions about the technological expertise of the civilizations being sought out[1,2,3]. A number of searches have looked for optical signatures, though few were able to be done systematically due to practical and funding limitations [4-13]. We will assume that the civilizations we are seeking have directed energy capability to equals or exceed our currently and reasonably projected capability in the near future. This is a modest assumption given the rapid advances in this area and we will see that we already possess the basic technology to see and be seen across the entire horizon. In particular we will assume that the civilizations possess the ability to build the equivalent of our DE-STAR program, namely phased arrays of lasers. This allows for a significant advances beyond what has previously been done and has the long term capability allowing extremely large systems. It is this latter that dramatically changes the SETI analysis. We assign the same civilization classifications (denoted as S) scheme as we use for the DE-STAR array classification where the civilization class indicates both the power level and beam size of the emitted laser. We assume a standard DE-STAR (S) with nominal Earth like solar illumination ($F_E = 1400$ w/m$^2$ at the top of the atmosphere) and a square laser array size (d) where $d(m)=10^S$ and beam divergence full angle $\theta = 2\lambda(m)/d(m) = 2\lambda 10^{-S}$ and solid angle $\Omega(st) = \theta^2 = 4\lambda^2 10^{-2S}$ for small angles. The power is assumed to be CW rather than pulsed with a value of approximate $P(kw) = 1.4\, \varepsilon_c\, 10^{2S}$ where $\varepsilon_c$ is the conversion efficiency of solar to laser power (eff$_{pv}$ * eff$_{de)}$.

The critical observable is the flux (w/m$^2$) at the (Earth) telescope and this is the transmit power P (w)/L$^2$ Ω where L(m) is the (luminosity) distance. Thus the critical ratio at given distance is P(w)/ Ω(st). For a DE-STAR system of class S we have
P(w)/ Ω(st) = $F_E$ $\varepsilon_c$ $10^{2S}$/4 $\lambda^2$ $10^{-2S}$ =1400 $\varepsilon_c$ $10^{2S}$/4 $\lambda^2$ $10^{-2S}$ = 350 $\varepsilon_c$ $\lambda^{-2}$ $10^{4S}$.
We can thus calculate the civilization class S from any system with a given power and solid angle, even if not a DE-STAR class system, as:

S = ¼ $\log_{10}$ ([P(w)/ Ω(st)]/( 350 $\varepsilon_c$ $\lambda^{-2}$)) = ¼ $\log_{10}$ ([P(w)/ Ω(st)]/( 175 $\lambda^{-2}$)).

We assume $\varepsilon_c$ = 0.5 total conversion efficiency of solar (stellar) illumination to laser output. This is about a factor of two higher than our current state of the art for CW systems (present efficiency of concentrated space solar is 50% and laser efficiency is above 50% for the most efficient systems).
For reference a class 0 civilization would possess the equivalent of a 1 meter diameter optical system transmitting approximately 1 kw while a class 4 civilization would be able to build a 10 km array with transmitting approximately 100 Gw and a class 11 civilization would be able to harness the power of a star like our Sun and convert it into directed energy. A class 5 civilization would be similar in this sense to a Kardashev Type I while a class 11 civilization would be similar in this sense to a Kardashev Type II or similar to civilization that can harness a typical star. We are currently about a class 1.5 civilization and rising rapidly. We already have the technological capability to rise to a class 4 civilization in this century should we choose to do so. As one example, two class 3 and above civilizations can "see" each other across the entire horizon modulo the time of flight. Here we use the term (entire horizon) to refer to high redshift galaxies we feel have had sufficient time to develop life. This is discussed further below.

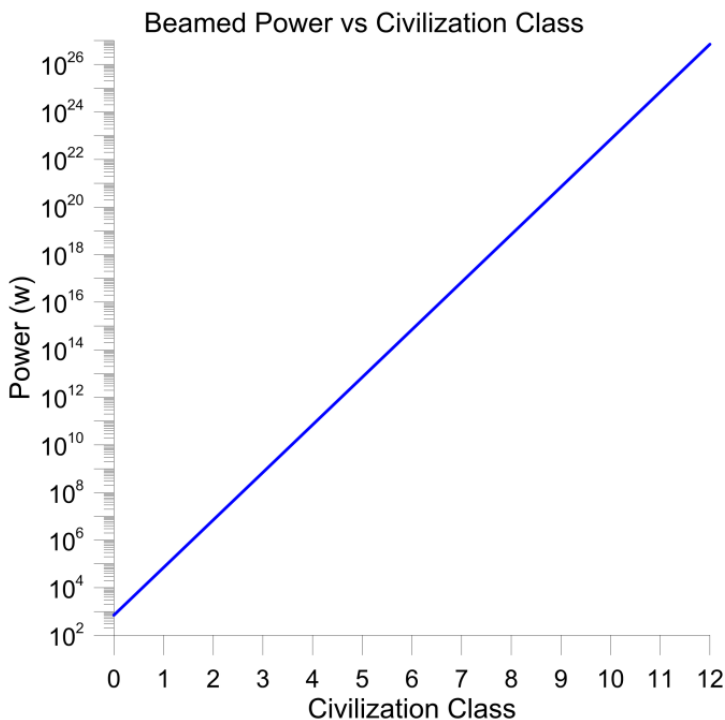

**Figure 2** – Civilization class and laser emitted power level (CW).

**Flux and Magnitude Equivalents vs Civilization Class and Distance** - We can now compute the flux at the Earth from a distant civilization which we show in Figure 3. The distances are the effective "luminosity distance" which at non cosmological distances is simply the normal Euclidean distance we are used to measuring. At cosmologically significant distances we need to use the cosmological correction reflecting the geometry of our universe. This is discussed and computed below. It is helpful to also think of the received flux in terms of the equivalent photometric magnitude that is commonly used in astronomy. We show this in Figure 4 as a rough indication of how "bright" the signal is. The equivalent magnitude is computed as if the signal were uniformly distributed over the typical photometric bandwidth of R~ 4. Of course the laser lines we look for are much narrower so we have vastly less background that in a photometric band. Nonetheless this is instructive when comparing to the common language of magnitudes in astronomy. **As can be seen at the distance of the typical Kepler planets (~ 1 kly distant) a class 4 civilization (operating near 1μm) appears as the equivalent of a mag~0 star (ie the brightest star in the Earth's nighttime sky), at 10 kly it would appear as about mag ~ 5, while the same civilization at the distance of the nearest large galaxy (Andromeda) would appear as the equivalent of a m~17 star. The former is easily seen with the naked eye (assuming the wavelength is in our detection band) while the latter is easily seen in a modest consumer level telescope.**

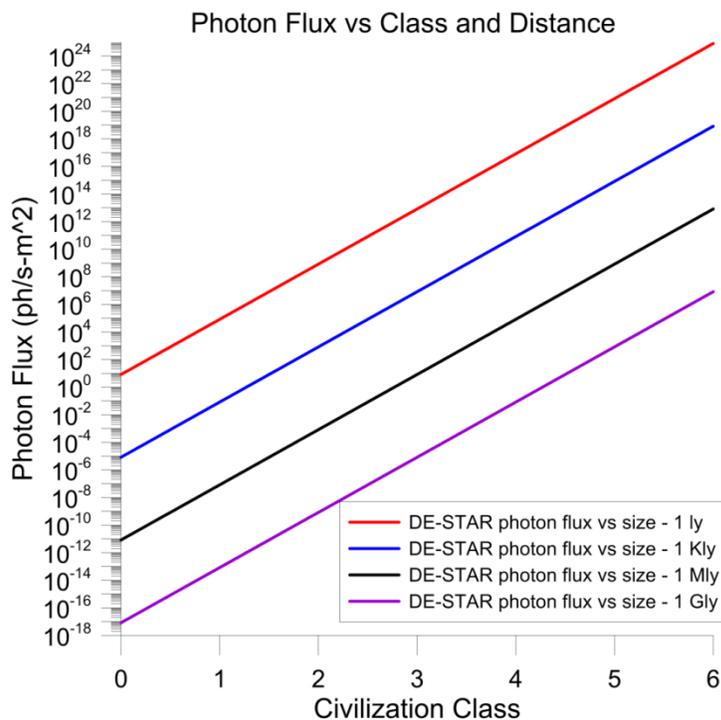

**Figure 3 –** Photon flux at Earth vs civilization class and distance. Distances are luminosity distance. See below for cosmological effects at higher redshift.

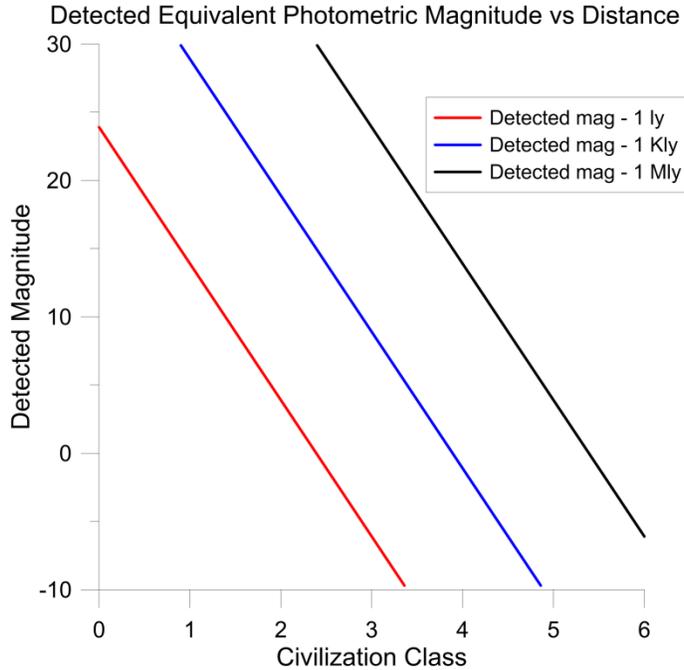

**Figure 4** – Equivalent photometric magnitude vs civilization class and luminosity distance. At distances small compared to cosmological scales the Euclidean distance and luminosity distance are equivalent. The equivalent photometric magnitude is based on an equivalent R~ 4 photometric filter band.

## 4. ATTENUATION AND GRAVITATIONAL LENSING

### 4.1 K Corrections due to dust and gas

Gas and dust in interstellar and intergalactic space absorb and scatter radiation. This is sometimes known as "reddening" since the SED from distant stars and galaxies is shifted towards the red portion of the spectrum as the dust preferentially absorbs and scatters the shorter wavelength light (the "bluer part") and allows more of the longer wavelength portion (the "redder portion" to pass through. This is analogous to the reddening of the sun at sunset. The details of this process depend on the form and distribution function of the dust grains. Normally objects are studied whose host spectrum is assumed to be known and the observed spectrum is a measure of the dust. The difference between the as observed and as emitted vs wavelength is known as the "K correction". K is conventionally given in magnitudes and depends on wavelength, direction of the target and distance to the target. It is also conventional to use a K correction to take account of the atmospheric transmission discussed below. In general the shorter wavelengths are absorbed more by dust and gas while the longer IR wavelengths are much less affected. The interaction with neutral gas is generally quite small except when the photon energies are above an ionization energy which is not the case in the IR except for very rare cases highly excited states. Ionized gas in the ISM and IGM is another source of interaction between photons and matter (primarily electrons here) but the densities on average are low enough that this is not a serious concern except in (rare) highly compact regions.

$F_0(\lambda)$ = flux without dust and gas

$F(\lambda)$ = flux with intervening dust and gas

$m_0(\lambda)$ = magnitude without dust and gas

$m(\lambda)$ = magnitude with intervening dust and gas

$\alpha(\lambda)$ = attenuation coefficient from dust and gas

$K(\lambda)$ = K correction magnitude due to intervening dust and gas

Note that $\alpha(\lambda)$ depends on the target direction and distance

$F(\lambda)/F_0(\lambda) = e^{-\alpha(\lambda)} \equiv$ transmission

Since magnitude differences are defined as the log of flux ratios we have:

$K(\lambda) \equiv m(\lambda) - m_0(\lambda) \equiv -2.5\log[F(\lambda)/F_0(\lambda)] = 2.5\log[e^{-\alpha(\lambda)}] = 2.5\alpha(\lambda)\log(e) \sim 1.086\alpha(\lambda)$

$m(\lambda) = m_0(\lambda) + K(\lambda)$ {hence the term K correction}

The transmission thru the dust and gas is given by:

$F(\lambda)/F_0(\lambda) = e^{-\alpha(\lambda)} = e^{-K(\lambda)/2.5\log(e)} \sim e^{-0.921K(\lambda)}$

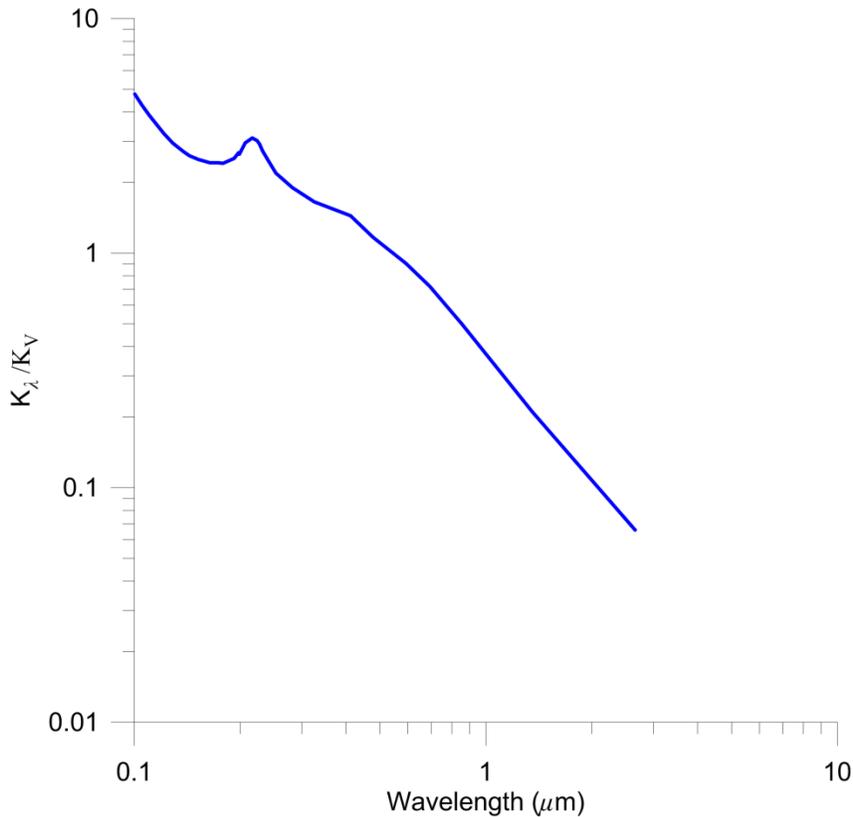

**Figure 5** - Ratio of extinction coefficient at a given wavelength to the same but in V band (~ 0.5 microns) in our galaxy. Note this is an approximation as the extinction coefficients are anisotropic. As is typical the extinction coefficient decreases with increasing wavelength.

## 4.2 Gravitational Lensing

Gravitational lensing occurs due to the gravitational interaction of photons with the gravitation field due to matter (both Baryonic and Dark). Gravitational lensing is well known but not on the small angular scales that may be relevant here. In addition there is a time varying component due to the motion of matter. There are numerous studies of gravitational lensing in the visible as well as the large scale power spectrum studied by the Planck mission (Planck collab 2016)[15]. The primary issue here is less the overall deflection of the beam but rather the gravitational focusing and defocusing that may occur on close approaches to stars (Maccone 2009)[16]. This overall area requires a more sophisticated simulation for various realizations and will not be covered in this paper.

## 5. FUNDAMENTAL BACKGROUNDS

### 5.1 Backgrounds relevant for detection

In order to determine the signal to noise of the return signature it is necessary to understand the non-signal related sources of photons. This is generically referred to as the background. There are a number of such backgrounds that are important. Going outward from the detector to the target and beyond, there is:

- Dark current and "readout noise" associated with the detector

- Thermally generated photons in the optical system, under the assumption that the optical system is mostly running near 300 K.

- Photon statistics of the received signal.

- Atmospheric emission – sky glow if the observations are inside the Earths atmosphere.

- Solar system dust that both scatters sunlight and emits from its thermal signature. Dust in the solar system is typically at a temperature of about 200 K. This is generically called Zodiacal scattering and emission, respectively, or simply Zodiacal light. This assumes a mission inside the solar system. We assume that there is a similar level of equivalent dust in the host civilization "solar system"

- Distant background stars that are in the field of view

- Sunlight scattered into the field of view for targets that are near to the sun in the field of view. This is generally only important for targets that are very close to the sun along the line of sight, though off axis response of the optical system can be an issue as well.

- Scattered galactic light from dust and gas in our galaxy.

- The far IR background of the universe, known as the Cosmic Infrared Background or CIB. This is the total sum of all galaxies (both seen and unseen) in the field of view in the laser band.

- The Cosmic Background Radiation or remnant radiation from the early universe. This is negligible for short wavelengths.

In all of these cases the fact that the laser linewidth (bandwidth) is extremely narrow (from kHz to GHz depending on the laser design) and the field of view is extremely narrow, mitigates these effects which would otherwise be overwhelming for a broadband photometric band survey. Heterodyning is also possible could be used in the future but is not assumed as we do not posses large focal plane arrays of such detectors.

## 5.2 Cosmic IR Background - CIB

The CIB was first detected by the Diffuse IR Background Explorer (DIRBE) instrument on the Cosmic Background Explorer (COBE) satellite launched in 1989 and studied by numerous other experiments including the recent Planck mission.[27,28,29,30] It is an extremely faint background now thought to be due to the sum of all galaxies in the universe from both the stellar (fusion) component at short wavelengths near 1 μm and from the re-radiated dust component near 100 μm. On large angular scales (degrees) it is largely isotropic though at very small angular scales (arc sec) individual sources can be detected. The diffuse CIB component, using data collected by DIRBE, is shown in Fig. 6.

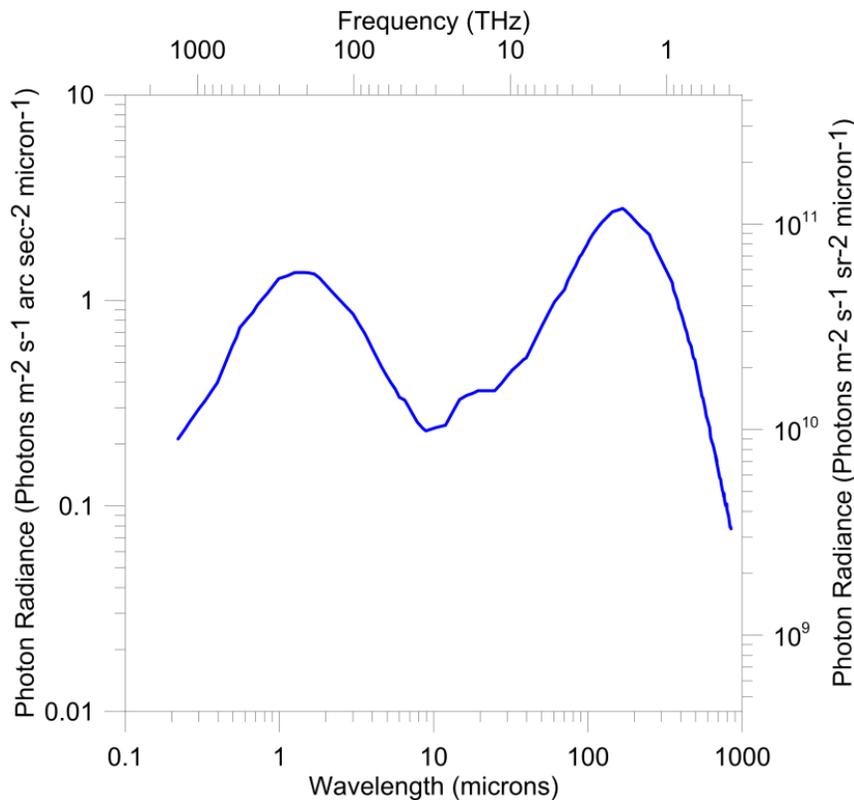

**Figure 6 -** Cosmic Infrared Background vs wavelength. Note the contribution from the stellar fusion peak near 1 micron and the reradiated dust peak near 100 microns.

## 5.3 Zodiacal Light

Like the CIB the zodiacal light has two components and both involve dust in the solar system and the Sun. The sunlight both scatters off the interplanetary dust grains giving a "streetlight in fog" effect as well as heating the dust grains which then reradiate in the mid to far IR. The scattered component can be seen with the unaided eye in dark extreme latitudes and is sometimes known as the "Gegenschein" and traces the ecliptic plane. The dust grains are in rough equilibrium through being heated by the Sun and cooling through their own radiation. This background is not isotropic but is highly anisotropic depending on the position and orientation of the observer in the ecliptic plane. This was studied in detail by the DIRBE instrument on COBE.[27,28,29] As seen in Fig.7, based on some of the DIRBE measurements, the brightness of both the scattered and emitted components vary dramatically with the observed line of sight relative to the ecliptic plane. In the plot the angle relative to the ecliptic plane is given by the ecliptic latitude (Elat) where Elat = 0 is looking in the plane and Elat = 90 is looking perpendicular. The situation is even more complex as the scattered and emitted components vary with the Earth's position in its orbit around the Sun. By comparing the CIB and the Zodiacal light, it is clear that even in the best lines of sight (perpendicular to the ecliptic plane) the Zodiacal light completely dominates over the CIB. For the JWST mission the Zodiacal light is typically the limiting factor for IR observations, for example. However, since illumination will occur in a system with an extremely narrow laser bandwidth, and detection occurs with a matched narrow bandwidth (allowing for Doppler shifting) , it is possible to largely reduce the Zodiacal light and the CIB to negligible levels. This is not generally true for broadband photometric (typically 30% bandwidth) surveys.

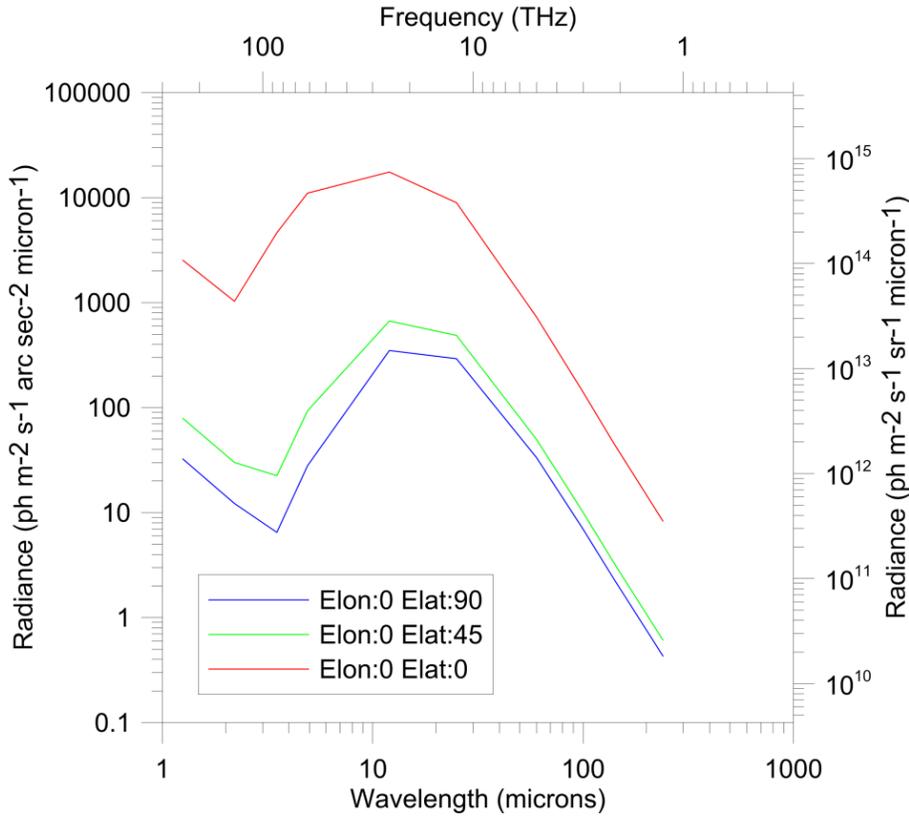

**Figure 7** - Zodiacal light emission vs wavelength and observing angle relative to the ecliptic plane. Note the reradiated dust peak near 10 microns.

### 5.4 Optics Emission

The optical emission from the telescope also needs to be considered. The optics are assumed to be at roughly 300 K for simplicity (this could be changed in some scenarios), giving a brightness of about $1\times10^7$ ph/s-m$^2$-sr-μm for unity emissivity (or for a blackbody emitter) at the baseline wavelength of 1.06 μm. Unity emissivity is clearly an over estimate but represents a worst case. Under the assumption of a diffraction limited system, the entendue of the optics is such that $A\,\Omega = \lambda^2 \sim 10^{-12}$ m$^2$·sr where A is the effective receiving area and $\Omega$ is the received solid angle. The bandwidth of reception must also be included. Here a matched filter spectrometer or heterodyning is assumed (to get Doppler) with a bandwidth equal to the laser linewidth. As mentioned above, this is typically $10^4$ - $10^{10}$ Hz or approximately $4\times10^{-11}$ to $4\times10^{-5}$ μm. The total per sub element is thus an emission of about $4\times10^{-16}$ to $4\times10^{-10}$ ph/s again for an emissivity of 1. This is an extremely small rate compared to the other backgrounds (air glow, Zodi, CIB) as well as the signal itself. Comparing the optics emission of $1\times10^7$ ph/s-m$^2$-sr-μm for unity emissivity to the CIB and Zodiacal light shows the CIB and Zodiacal light are both much larger than the optics emission.

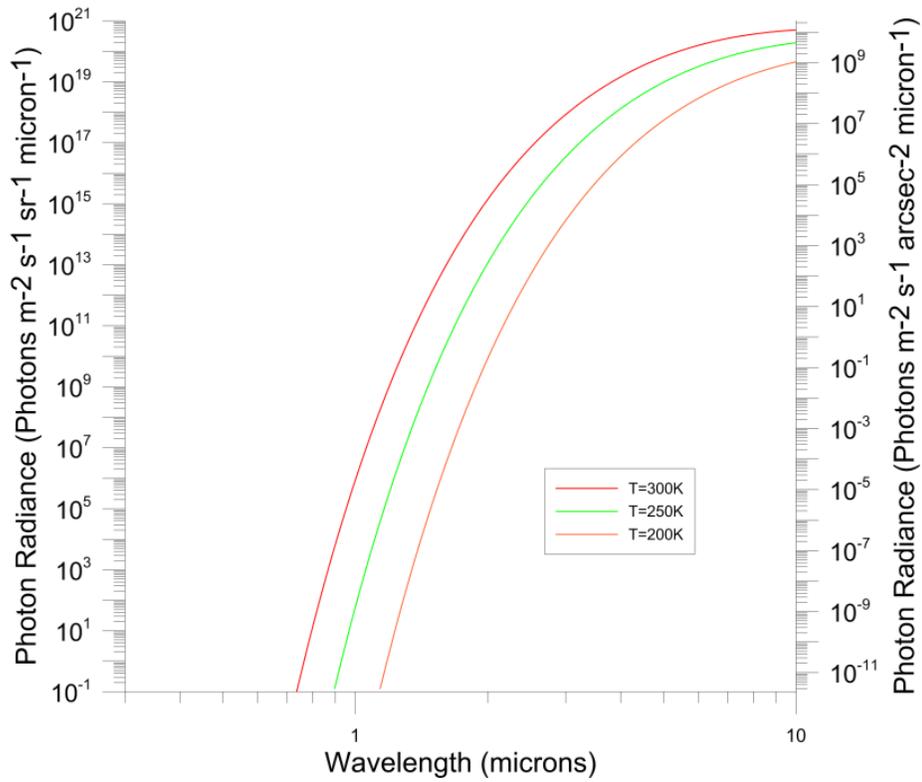
**Figure 8 -** Optical emission assuming unity emissivity.

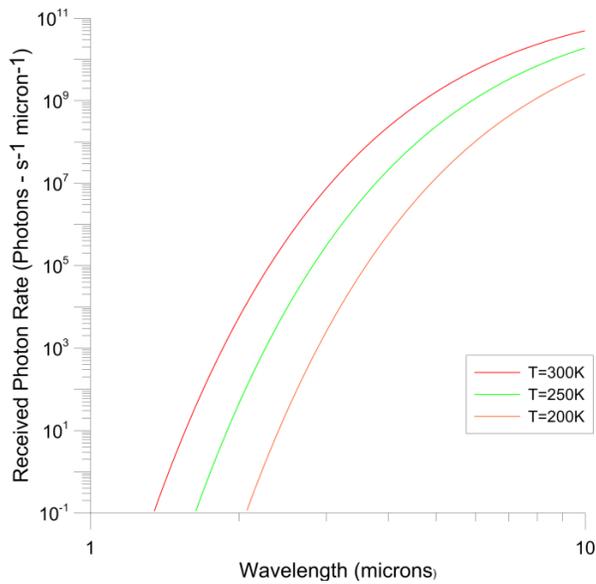
**Figure 9 -** Optical emission assuming a diffraction limited optical system.

## 5.5 Atmospheric Transmission and Radiance

For studies inside the Earth's atmosphere we need to consider the transmission and emission of the atmosphere. We consider the transmission and thermal radiance of the Earth's atmosphere for different observation scenarios from sea level, to high mountain observatories to aircraft and finally stratospheric balloons. There are a number of observational windows that allow us to observe in the visible and IR that must be taken into account to optimize a search strategy especially one at high redshift. We will see that observations at high redshift become feasible for some scenarios. In addition to atmospheric thermal radiance we consider non thermal processes below as well as anthropomorphic produced lines.

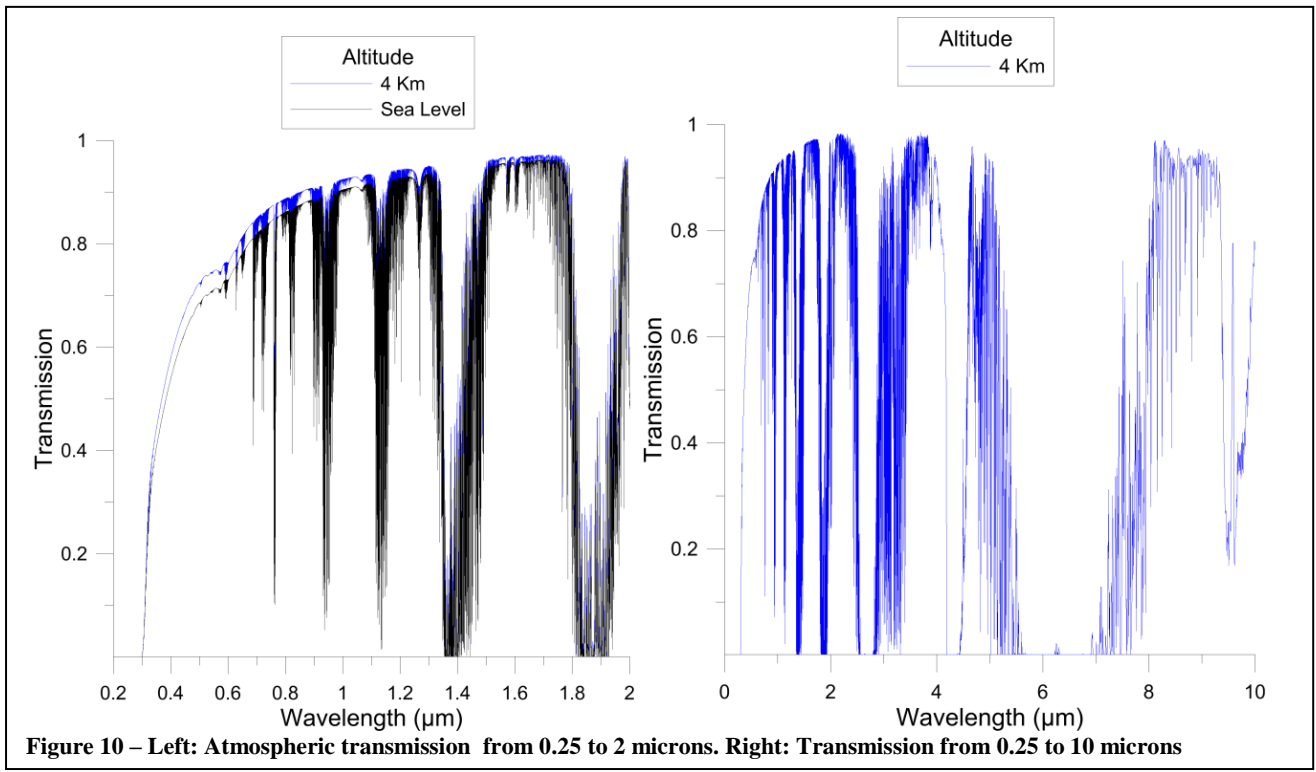

**Figure 10 – Left: Atmospheric transmission from 0.25 to 2 microns. Right: Transmission from 0.25 to 10 microns**

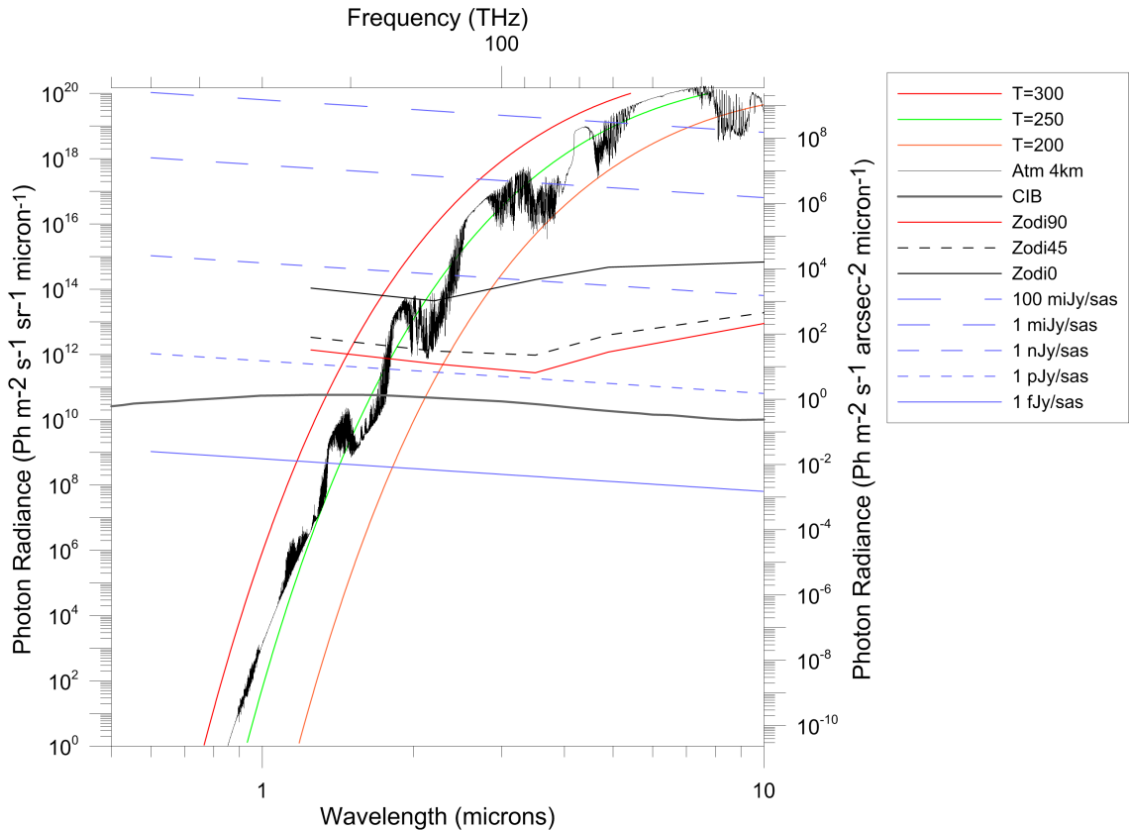

**Figure 11 -** Thermal emission from optics, atmospheric thermal radiance, CIB and Zodiacal light in the ecliptic plane (0) at 45 degrees relative to the plane (45) and perpendicular to the plane (90). Zodi is for COBE DIRBE day 100.

## 5.6 Non LTE Atmospheric Emission

There are additional processes in the Earth's atmosphere that are not in local thermodynamic equilibrium with the atmosphere. In particular various atomic and molecular transitions are excited by the solar wind and other energetic phenomenon. In the visible and IR there are a variety of non LTE lines that are highly time variable include Oxygen and OH emission. In general these have modest low spatial frequency variations but the variable background rates will be an issue at extremely low intensities. OH emission originates at altitudes above 80km typically and is most problematic in J (1.1-1.3 microns) and H (1.5-1.8 microns) bands with some in K (2-2-4 microns) band. Rousellot et al (2000) have computed the theoretical OH spectra of 4732 lines from 0.6 to 2.6 microns and spectrometers at major telescope measure the brighter OH lines. As mentioned the OH line emission is highly variably both temporally and spatially. OH lines are extremely narrow (unresolved at R=10,000 where R=$\lambda/\Delta\lambda$) and while there are many lines they occupy a very small fraction of the spectrum due to their narrow linewidth. There is also a very large dynamic range in predicted OH line emission (over 14 orders of magnitude). Only the brighter lines are typically visible and longward of 2.6 and shortward of 0.6 there is very little OH emission. We also show a zoom in near the 1.064 micron Yb transition that is the baseline for our larger DE-STAR system as an example of the narrow nature of the lines and their spacing near the Yb line. This is one example. Fortunately we can achieve some additional rejection of OH due to the assumed point like structure of the source we are looking for while OH is spatially broad so some spatial filtering will be useful.

This is analogous to photometry determination of the local sky background in aperture photometry. Comparing OH emission in J and H bands it is clear that the OH lines dominate when using broad band filters while in the visible bands and beyond K band OH lines become sub dominant. This applies to ground based measurements while for space based measurements OH lines are not relevant. Since the OH lines are very narrow reducing the filter bandwidth does not allow us to completely mitigate them until we get to extremely narrow band filters or use an IFU both of which are problematic. Note that in a filter bandwidth the total OH emission is the sum of all the OH lines within the band. The use of aperture photometry and synthetic sky techniques will help us model and reduce the effects of OH line emission (as with all large angular scale emission) but we are still left with the noise from both photon statistics and systematic errors that will need to be taken into account. In this sense the problem is similar to classical LTE emission from optics and the atmosphere as well as from the detector but with the added complexity of more challenging temporal and spatial variations in the OH emission. In the visible bands and beyond 2.4 microns the OH emission is relatively small. The primary problem occurs between 1 and 2.4 microns. For broad band photometric systems non thermal emission dominate out to about 2 microns. For narrow bandwidth or spectroscopic systems zodiacal emission and scattering dominates out to about 1.5 microns. In the long run space based searches are preferred.

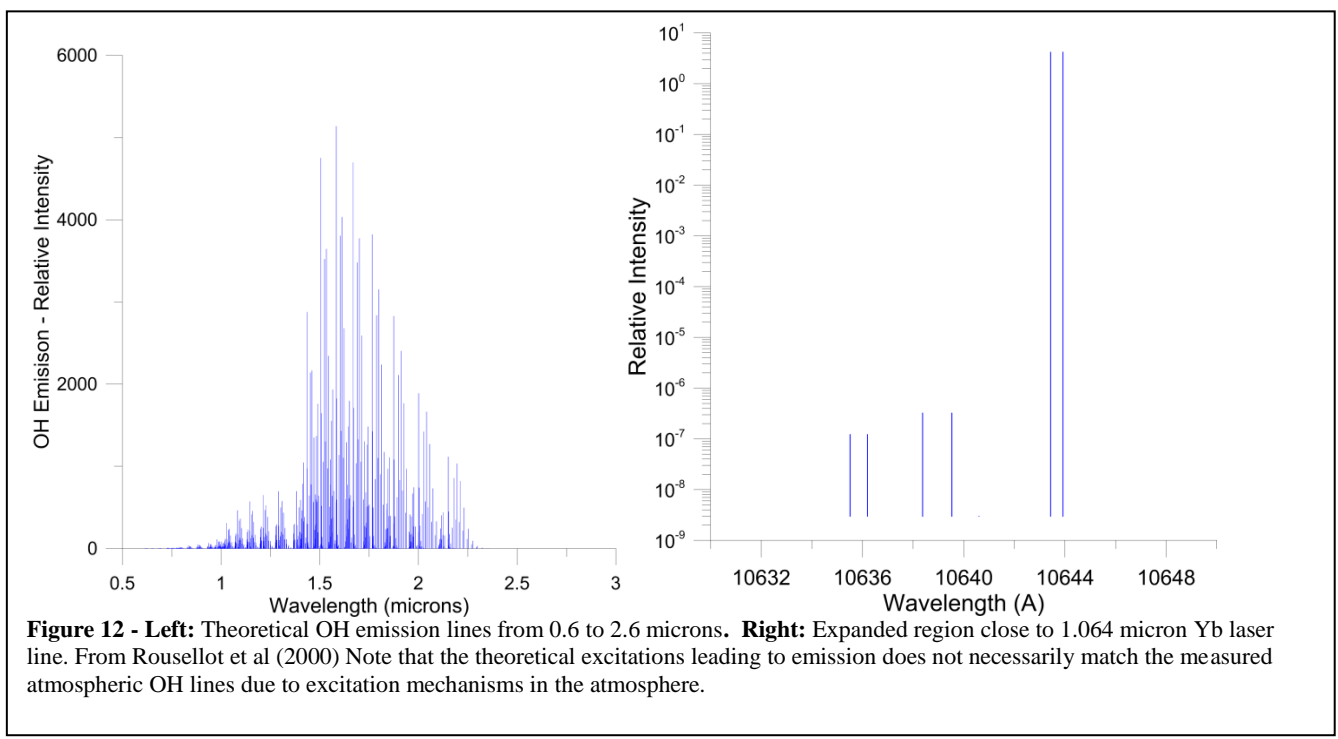

**Figure 12 - Left:** Theoretical OH emission lines from 0.6 to 2.6 microns. **Right:** Expanded region close to 1.064 micron Yb laser line. From Rousellot et al (2000) Note that the theoretical excitations leading to emission does not necessarily match the measured atmospheric OH lines due to excitation mechanisms in the atmosphere.

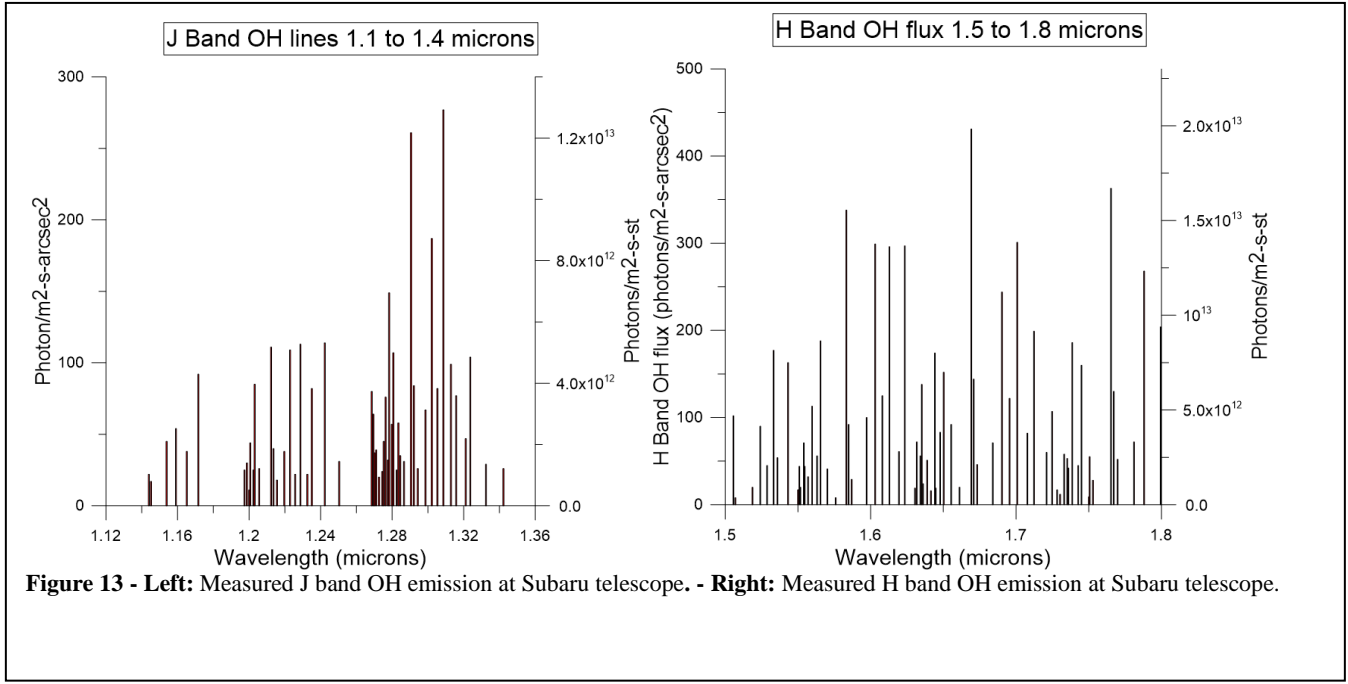

**Figure 13 - Left:** Measured J band OH emission at Subaru telescope. **- Right:** Measured H band OH emission at Subaru telescope.

## 5.7 Measured Total Sky Background

For the best observatory sites the sky background minimum is about 21-22 mag/sq arc sec in V band (centered at $\lambda \sim 0.55\mu m$ with bandwidth $\Delta\lambda \sim 0.1\mu m$). This corresponds to a flux of approximately 10-50 photons/s-m$^2$ -sq arc sec. This includes thermal as well as non thermal processes (air glow), zodi, unresolved stars etc. Comparing to the figures above we see this is in reasonable agreement. In the V band the dominant emission is from Zodi scattering of sunlight as well as non thermal atmospheric (air glow) processes. As we move towards into IR the thermal emission of the atmosphere and optics as well as OH lines begin to dominate with OH diminishing beyond K band (2.4 µm).

## 5.8 Terrestrial illumination

Human lighting is an issue but in general is not as severe for our search as it tends to be a relatively slow temporal and spatial function. Some Hg and Na lines from HID lights are notable and increasingly LED lighting though the latter is generally broadband due to phosphor coatings. All of these are site dependent and can be mitigated by observing targets at multiple locations and over multiple time scales.

## 5.9 Stellar and Interstellar line emission

Host and intervening stellar atmospheres will provide some confusion due to the emission lines and to a lesser extend from absorption lines. In addition to common know lines we can also check their temporal distribution to see if they are natural or not. Using temporal photon statistics allows us an additional cross check as well as more conventional tests for unnatural time modulation of possible positive targets.

## 5.10 Unresolved stellar background

In many surveys we will not resolve individual stars and thus will have many stars per pixel. These unresolved stars will form a background, much like the CIB. Since the stellar distribution in galaxies is a strong function of position in the galaxy it is unlike the CIB in this sense and is highly spatially variable. This has implications for the coupling of pointing jitter and seeing variations into our data. In particular the unresolved stars have emission lines that will form a line background in addition to the continuum background. For example the dark sky background of 22 mag/sq arc sec in V band includes the unresolved stellar background among other backgrounds. As one example consider stars like our Sun. The Sun has an absolute magnitude (apparent magnitude if it were placed at a distance of 10 pc) of $M_v = 4.83$ and an apparent magnitude $m_v$ vs distance d(pc) of $m_v(d) = M-5+5\log(d(pc))$. Imagine we place the Sun at 10 kpc (approximately the distance from Earth to the galactic center and about 1/3 the "diameter of our galaxy). The apparent magnitude of our Sun would then be $m_v(d=10 \text{ kpc}) = 19.8$. To put this in perspective the photon flux of $m_v=0$ star is about $10^{10}$ γ/s-m$^2$ (this depends on the equivalent temperature of the star). Hence a star with $m_v=20$ (approx that of our Sun at d=10 kpc) would have a flux of 100 γ/s-m$^2$. Our galaxy has an average stellar density of approximately 1 star/sq arc sec. If our galaxy has a uniform distribution of stars like our Sun all at a distance of 10 kpc then would expect a stellar flux of about 100 γ/s-m$^2$-sq arc sec in V band which is close to the dark sky flux of about 10-50 γ/s-m$^2$-μ-sq arc sec in V band. Since the flux from a laser associated with a planet near a star and the flux from the parent star both scale inversely with the square of the distance to the star we will see that the stellar flux is a relatively small noise source when we calculate signal to noise ratios. Note that for a diffraction limited system (1 pol) $A\Omega=\lambda^2$. As the unresolved stellar background signal is proportion to both the telescope area and solid angle (per pixel) the total signal for the diffraction limited case is independent of the telescope size.

## 5.11 Unresolved galactic signatures

Along any given direction we will have a number of distant galaxies in a pixel for ground based surveys as well as small aperture space based surveys. This is basically the CIB but in this instance there is an additional component to the usual CIB in that each galaxy has some billion to trillion possible civilizations. On average in a square arc second, typical of ground based seeing without adaptive optics, we will have an unresolved and undetected distant galaxy at an unknown redshift.

## 5.12 High Redshift surveys

We can detect civilizations at a variety of redshifts and this poses unique opportunities and challenges. For higher civilization classes we can detect them at any redshift which is both good and problematic for our detection algorithm.

We show the relationship between distance and redshift in the attached plot for several cosmological models. There is relatively little difference in the models for luminosity distance even ignoring dark energy at low redshift. The distances we normally quote are luminosity distances even if just labeled distance. We also show cosmological age vs redshift and cosmological age vs luminosity distance. By redshift z=5 the age of the universe is only about 1.2 Gyr. If life does not evolve rapidly after star formation then there would not be sufficient time to evolve technologically advanced civilizations capable of emitting detectable directed energy signatures. The luminosity distance at z=5 is about 47 Gpc corresponding to a Euclidean distance of about 150 Gly. While still detectable for some higher civilization classes the time for advanced technological evolution is short. Correspondingly at z=1 the cosmological age is about 5.8 Gyr corresponding to a luminosity distance of about 6.7 Gpc allowing much more time for life to evolve. For reference our evolution on Earth is about 3-4 Gyr. We also show the comoving volume of the universe vs the redshift we observe to as well as the normalized comoving volume explored to a given z relation to z=20 where we chose z=20 to be a reasonable approximation for the first stars and planets. Note that z=20 contains the vast majority of the volume of our horizon but that z=20 is only about 150 Myr after the beginning and this is likely not sufficient time for intelligent life to form. If we assume intelligent life needs 4.5 Gyr to form (approximately our evolution time after the formation of the solar system) this would correspond to about z~1.5. We also show the normalized comoving volume normalized to z=1.5. The normalized comoving volume is essentially the fraction of the accessible universe where we might expect to find technologically advanced life based on our own evolution. These are obviously large assumptions on our part. We use a concordance model (2015 Planck) which yields a current age of around 13.8 Gyr.

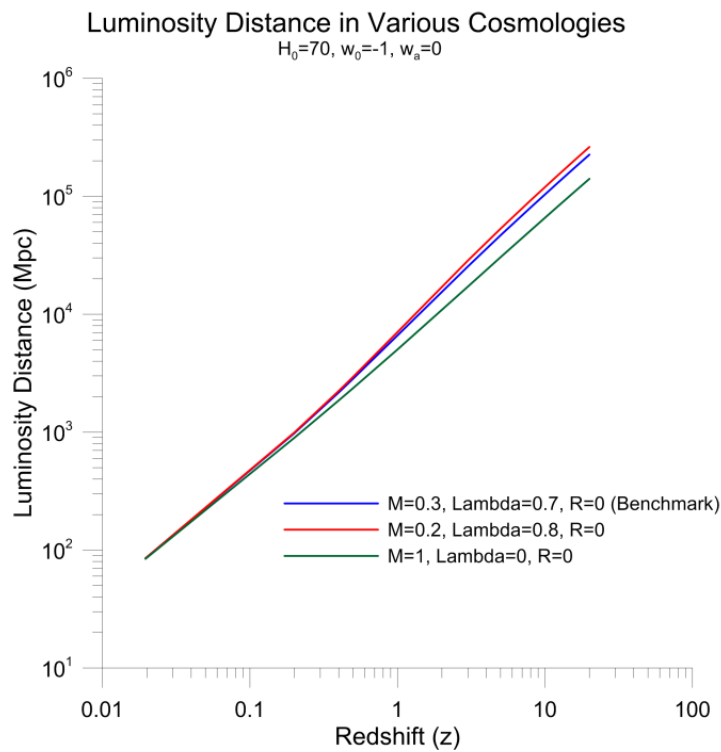

**Figure 14 -** Luminosity vs Redshift for several cosmological models. The "benchmark" model is closest to the current concordance models. This is used in the calculations below for higher redshift models.

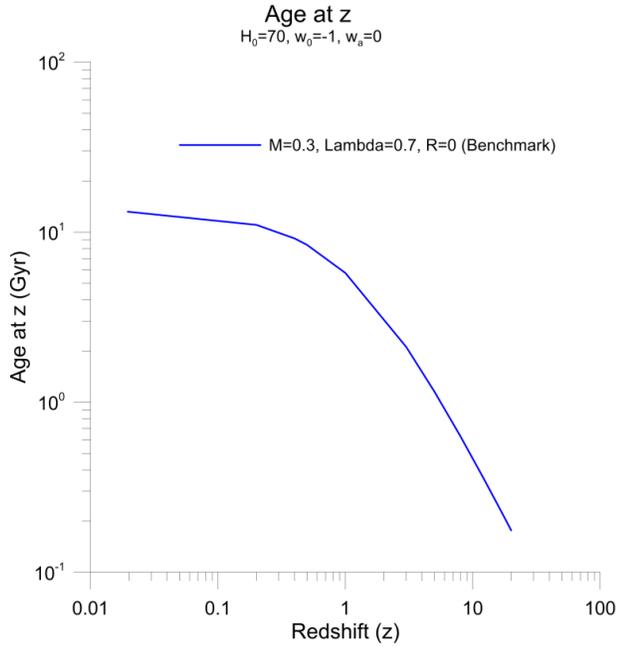

**Figure 15 -** Age of the universe vs redshift for the current concordance model. This is critical for understanding the possibilities of life forming in enough time at high redshift. Concordance universe model used.

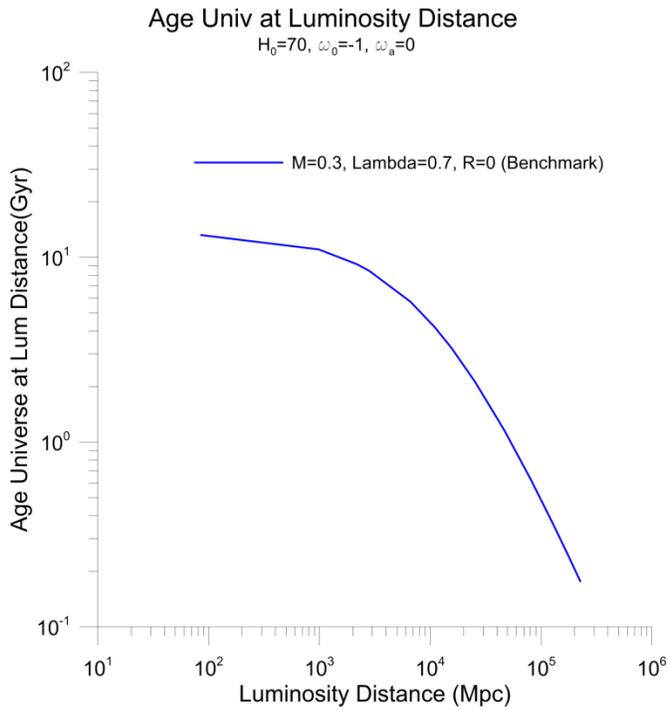

**Figure 16 -** Age of the universe vs luminosity distance. This is used in the discussion of the time scale for the evolution of life and the effective luminosity distance it corresponds to. Concordance universe model used.

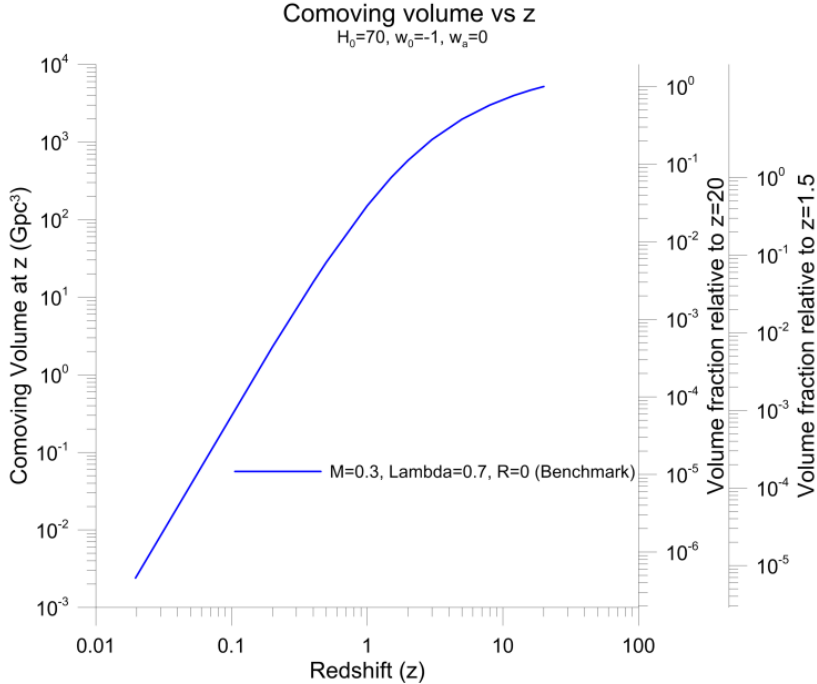

**Figure 17 -** Comoving volume vs redhsift. Also shown is the normalized fraction of the volume at z=1.5 and 20. By z=20 virtually all the comoving volume is explored while at z=1.5 a bit less than 10% of the volume is. Concordance universe model used.

## 5.13 Detection bandwidth

The intrinsic bandwidth of lasers is extremely narrow by most astronomical standards. Laser lines as narrow as 1 Hz or even less have been demonstrated. For current high power laser amplifiers the bandwidth is typically at the 0.1-1 KHz level but to achieve the highest power levels this is artificially broadened to about 10 GHz/Kw , at a wavelength near 1 micron,  to overcome the Stimulated Brillion Scattering (SBS) limits in the fibers. There is no intrinsic reason this broadening needs to be implemented but is done due to current technological limitations.  Indeed, if lower power per fiber amplifiers are used and indeed at the 10-100 watt amplifier level bandwidths below 1 KHz are already achievable. The bandwidth language of lasers is usually given in Hz while the astronomical language of bandwidth is usually discussed in microns or nanometer. The relationship between the two is simply $\Delta\lambda = c\nu^{-2} \Delta\nu$. The effective spectroscopic resolution is defined as $R = \lambda / \Delta\lambda = \nu / \Delta\nu$. To put this in perspective a laser line at 1 micron ($\nu \sim$ 300 THz) with a 1 Hz bandwidth (mixing units is typical in this field unfortunately) has an $R \sim 3 \times 10^{14}$. By astronomical standards of spectrometers this is a phenomenally large R. Even with the current broadened SBS limit mitigation techniques of 10 GHz/Kw the effective $R \sim 30{,}000$ for a 1 Kw fiber amplifier at 1 micron. The current state of the art for astronomical spectrometers whether fiber fed or free space is about $10^4$ - $10^5$. Heterodyne spectroscopy is now becoming possible at optical and IR wavelengths and offers much higher R for the future if needed.

In the accompanying plot we show the laser linewidth (usually quoted in Hz) to the equivalent width in microns. We have chosen a wavelength of 1.06 microns for convenience. It corresponds to a particularly efficient Yb transition we are using as the baseline for the DE-STAR program but it is representative of any system. In this case the bandwidth in microns also corresponds to the equivalent $\beta = v/c$. While sources and receivers are in relative motion the effect is to shift the central line not to broaden it since the systems we envision are localized. The bandwidth is also approximately 1/R where R is the spectral resolving power for a 1 μm signal.

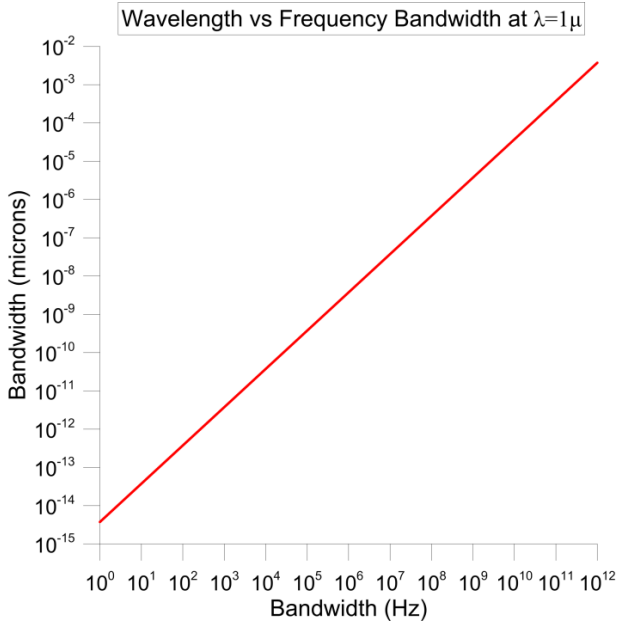

**Figure 18** - Bandwidth of laser line in microns vs Hz. Typically laser linewidths are specified in Hz while the more relevant parameter for astronomical discussion is in microns.

### 5.14 Comparison to in-band emission from natural sources

Since the laser line is very narrow it is important to understand how the "in-band" received flux integrated over the bandwidth of the line from the laser compares to natural sources of radiation. It is useful to compare the radiance (w/m$^2$-st) for a laser and for a star when both are integrated over their respective areas and over the linewidth or filter bandwidth. In this way we compare to emitters at their source assuming the laser is associated with a planet near a star.

For simplicity we model the star as a thermal source.

The brightness is then $B_\lambda = \frac{2hc^2}{\lambda^5 \left(e^{hc/\lambda kT}-1\right)}$ (W/(m$^2$-sr-m)). We then integrate this over the forward facing hemisphere of the star and over the linewidth and compare to the laser for a given civilization class. As an example we compare the brightness of our Sun, modeled as a 5700 K blackbody, at a wavelength of 1.06 microns and get
9 MW/m$^2$/sr-μm. With a diameter of 1.4x10$^9$ m this gives 1.4x10$^{25}$ w/sr- μm. The relationship between wavelength and frequency bandwidth Δλ=cv$^{-2}$ Δν at 1.06 microns is Δλ(μm)=4x10$^{-15}$ Δν (Hz). Assuming a laser linewidth of 1 kHz (typical for current state of the art modest power (~ 0.1 KW) amplifiers) this would yield Δλ(μm)=4x10$^{-12}$. For current high power amplifiers (KW class) that are SBS limit artificially broadened with a linewidth of 10$^{10}$ Hz this yields Δλ(μm)=4x10$^{-5}$. Both of these are small linewidths by astronomical standards but not by laser standards. Both linewidths currently exist in the relevant technology. However the primary effect of directed energy is in fact that it is directed. For example a class S civilization has a laser array size (d) where d(m)=10$^S$ and beam divergence full angle θ = 2 λ/d with a projected solid angle of approximately $\Omega = (2\lambda/d)^2 = 4\lambda^2 10^{-2S}$. As an example a class 4 civilization projects a beam with a solid angle of approximate $4 \times 10^{-20}$ sr.

1)- Comparing the Sun's power in the equivalent linewidth to a class 4 civilization gives the following power transmitted into the same solid angle :
Linewidth (1 KHz ~ $\Delta\lambda(\mu m)=4\times10^{-12}$)   Sun: $2\times10^{-6}$ W   Laser: $7\times10^{10}$ W
Linewidth (10 GHz ~ $\Delta\lambda(\mu m)=4\times10^{-5}$)   Sun: 20 W   Laser: $7\times10^{10}$ W

It is clear that the stellar light in a narrow bandwidth is very small by comparison to the DE source.

2)- For the resolved diffraction limited case (single mode and independent of array size and target distance IF resolved) we have A $\Omega = \lambda^2 = 10^{-12}$ m$^2$-st for $\lambda=1\mu$ which gives $9\times10^{-6}$ w/$\mu$m. The resolved case **only** applies for a large receiving array with a very nearby star and is generally not of interest as we do not think life is associated with the itself. With a linewidth of 1 KHz ($\Delta\lambda(\mu m)=4\times10^{-12}$) this gives $4\times10^{-17}$ w and $4\times10^{-10}$ w for a linewidth of 10 GHz.

In case 1) we treat the Sun as a prototype for a distant star, one that is unresolved in our telescope (due to seeing or diffraction limits) but one where the stellar light ends up in ~ one pixel of our detector. Clearly the laser is vastly brighter in this sense. Indeed for the narrower linewidth the laser is much brighter than an entire galaxy in this sense. *For very narrow linewidth lasers (~ 1 Hz) the laser can be nearly as bright as the sum of all stars in the universe within the linewidth.* **Even modest directed energy systems can stand out as the brightest objects in the universe within the laser linewidth.**

## 5.15 Orbital considerations and optimal detection bandwidth

As we do not apriori know the orbital speeds of the targets we are searching for we need to consider the optimum search strategy. There is also the issue of the bulk speed of the galaxy the target is embedded in. The shorter term, but predictable, orbital velocity variations due to the rotations of the Earth, orbit of the Earth around the Sun etc and the similar but unknown orbital environment of the target leads to a complex search optimization. Ideally broadband FFT like heterodyne searches will be possible in the future but we will concentrate on more (currently) practical methods such as using narrow band filters and IFU's. For example, if we adopt a series of narrow band filters as one approach to detection then one of the fundamental techniques is to temporally "chop" the spectral bands so that the Doppler shift due to orbital shifts during the period of observing for the shift over this period is small compared to the filter bandwidth. For example the earth's rotation speed (~ 1 km/s) yields a Doppler shift of roughly $3\times10^{-6}$ while the Earth's revolution around the Sun (~ 30 km/s) gives a Doppler shift of about $10^{-4}$. The time scales of these is very different being 1 day and 1 year respectively. Our typically observing times for a complete series of filters will be typically measured in hours so spectral chop period is even less than one rotation of the Earth. The equivalent for the target is completely unknown but we assume a comparable situation both for simplicity and based on the issue of habitability and known detected exo-planets.

## 5.16 Detection Bandwidth and Background Noise levels

To achieve the maximum signal to noise ratio we need to understand the level of the background noise vs the signal. The optimal filter bandwidth would be large enough to encompass the emitted laser line and any broadening mechanisms but not so wide as to significantly increase the background noise. On the other hand the filter bandwidth affects the search strategy if individual filters are used. There is a tradeoff. Ideally a large portion of the both spectral and spatial space would be simultaneous sampled to give a fast mapping speed. Currently there are no simple technical solutions consistent with both of these needs. Spectrometers exists with high R but they are limited to a very modest number of pixels. Our initial search strategy will focus of trading large simultaneous spatial coverage for large simultaneous spectral coverage. Ideally in the future this will change. Given all the sources of noise there is a point where having a filter that is too narrow becomes counter productive. This is the optimization that is required. For example if the filter bandwidth reduces the background levels to be much less than readout noise in the detector than no additional gain is added by reducing the filter bandwidth. There is a large parameter space to tradeoff here and with the target unknown it is simply a subjective trade. Adding in practical considerations such as telescope time and systems costs pushes the trade to larger filter bandwidths currently. Multichroic beam splitters is another option to increase thruput that we are exploring in addition to other techniques.

## 5.17 Search Strategies

To decide on a search strategy we first need to decide what it is we are looking for. At first this seems obvious (find "unnatural" sources, but the optimum search given limited time and resources is more subtle.

**Modulation detection** - One method is to look for sources of temporal or spatial modulation that is unnatural. If we focus on temporal modulation we think of laser communication modulation. For use this is typically in the Gbps or nanosecond modulation range. But this is another "anthropomorphic now" mindset. If we are observing at a wavelength around 1 micron the available bandwidth far exceeds 300 Tbps with proper encoding. We do not currently possess this technology nor is it obvious that given the time of flight for distances that are astronomically relevant that directed energy based data communication (streaming of "intelligent" information) would be logical. As always "we do what we can do". Hence searches for high frequency modulation at the reasonable limits of our current technology does make sense. There is of course no particular reason why civilization far more advanced that us would be transmitting data in the realm we can detect unless they are specifically trying to beacon other civilizations. This rapidly degenerates into a nearly useless philosophy of the unknown. See section below on blind beacons and searches for further discussion.

**Ignoring modulation and searching for narrow but unnatural lines** – **Massively parallel search strategies** - Another search option and the baseline we adopt is to search for narrow line emission that are unnatural and then follow up to determine if these lines contain intelligent information. The advantage here is we do not depend solely on temporal modulation to search but can observe extremely large number of possible targets simultaneously without knowledge of their modulation. In the past several groups have looked for short pulses as an indicator of unnatural sources. The advantage in this strategy is that high peak power can be produced much more easily in short pulses but the disadvantage is the average power of terrestrial pulsed lasers is generally significantly lower than CW systems and from a search strategy we have no apriori knowledge that extra-terrestrial civilization would use pulsed systems. Indeed on Earth we do not generally use pulsed systems for communications, though this should not be a guide. For a given amount of average power the SNR is not necessarily higher for a pulsed system. Another and much greater disadvantage to searching for pulsed signals is that they are usually done with a single pixel on the sky while in the CW search one can use a large format array detector with multiple megapixels (Gigapixels are possible now) and thus there is a tremendous parallel advantage to a CW imaging search. Both approaches should be used.

For example in any square arc second of the sky there is approximately one galaxy even if not currently known and in this galaxy there are approximately 100 billion solar systems IF the galaxy is similar in star and planet formation to our own. In a single square degree that are thus about $10^7$ galaxies and some $10^{18}$ possible stellar systems. This allows a massively parallel search strategy with no apriori pointing knowledge though we can directly image nearby galaxies. The fundamental issue here is to understand the SED (both line and continuum) well enough to model and subtract it. This then gets to the optimization of the filters. As we will see below, even with modest Earth based telescope, we can detect some advanced civilization across the entire horizon with current telescopes.

**Sources not directly beamed towards us** – A possibility is that we will "eavesdrop" on a laser communications system that is not intentionally beamed for other civilization detection. One option here is accidental line up (glint) that we just happen to intercept. The other option is to detect the side lobes or possible scattering of the main beam. The basic problem with these latter two is that the signal we would intercept would be drastically reduced as the typical side lobes and interstellar scattering is generally extremely small with far off axis side lobes reduced by a factor of $10^4$ to $10^{10}$ from the main lobe not being unusual. Scattering of the target of the "laser communications" system is another possibility but this also drastically reduces the observed flux.

## 5.18 Life at High Redshift

Life on Earth is thought to have evolved between 3 and 4 billion years ago with what we now call intelligent life being relatively recent. This puts the beginnings of life at about 1 billion years after the formation of the Earth. We have little idea of the "why" of the evolutionary path that life took and much was externally influenced by bombardments for example. The first stars in the universe are thought to have formed within a few hundred million years after the beginning of the universe. Planet formation presumably was on a similar time scale though the processes needed for life may have taken significantly longer. The times scales are sufficiently uncertain that we cannot rule out life at high redshift [14] and this would allow many billions of years more for life to evolve than on Earth. Our own technological capabilities are an extremely non linear function of time with virtually no technology being achieved until the 1 part per million. This places us on an extremely nascent portion of the curve of intelligence and technology. If we imagine not a few thousand years of technological evolution but a few billion years of this it becomes sobering to contemplate intelligent life evolving at high redshifts and having billions of years (or a million time more than our technological time scale) to grow technologically. As we look at any patch of sky, with a typical square degree field of view, we will be observing some 10 million galaxies or some $10^{17} - 10^{18}$ possible planets if high redshift planet fractions are similar to today. As shown below we can detect class 4 and above civilizations at high redshift even with modest ground based (meter class) telescopes if they transmit in our direction when we are observing.

# 6. DETECTION AND SIGNAL TO NOISE CALCULATIONS

We model the detection system in a standard way assuming a model for quantum efficiency, dark current, read noise, combined "sky background" etc.

$F$ = flux from target ($\gamma$/s-m$^2$)
$F_B$ = flux per solid angle from all background sources integrated over bandwidth ($\gamma$/s-m$^2$-st)
$B_B$ = flux per solid angle from all background sources per bandwidth ($\gamma$/s-m$^2$-st-$\mu$)
$A$ = telescope area (m$^2$)
$A_\varepsilon$ = effective telescope area including transfer efficiency and quantum eff = $A * \varepsilon * Q_e$ (m$^2$ e$^-$/ $\gamma$)
$\varepsilon$ = telescope transfer efficiency
$i_{DC}$ = detector dark current (e$^-$/s)
$Q_e(\lambda)$ = quantum efficiency of detector (e$^-$/ $\gamma$)
$\Omega$ = solid angle of pixel (st)
$\tau$ = integration time (s)
$S$ = signal due to source at detector over integration time (e$^-$)
$S_{DC}$ = signal due to dark current over integration time (e$^-$)
$S_B$ = signal due to background over integration time (e$^-$)
$S_{time}$ = total signal over integration time (e$^-$)
$N_R$ = readout noise (e$^-$)
$N_S$ = noise due to signal (shot noise) (e$^-$)
$N_{DC}$ = noise due to dark current (e$^-$)
$N_B$ = noise due to background sources (e$^-$)
$N_{time}$ = time dependent part of noise (not including readout noise) (e$^-$)
$n_t$ = signal, dark current and background noise (#e/Hz$^{1/2}$)
$N_T$ = total noise including read noise (e$^-$)
$S_N$ = signal to noise ratio = SNR = $S / N_T$

$$S = F \tau A \epsilon Q_e$$

(we assume we have dark field and bias subtracted the image)

$$S_{time} = S + S_{DC} + S_\beta$$

$$N_S = \sqrt{S} \qquad N_{DC} = \sqrt{S_{DC}} \qquad N_\beta = \sqrt{S_\beta}$$

$$N_{time} = \sqrt{N_S^2 + N_{DC}^2 + N_\beta^2} = \sqrt{S + S_{DC} + S_\beta} = \left(\tau(FA\epsilon Q_e + i_{DC} + F_\beta A\epsilon Q_e \Omega)\right)^{1/2} = \left(\tau n_t^2\right)^{1/2} = \tau^{1/2} n_t$$

$$n_t = \left(FA\epsilon Q_e + i_{DC} + F_\beta A\epsilon Q_e \Omega\right)^{1/2} \text{ Note units of are } \#e^- - s^{1/2} \text{ or } \#e^- / Hz^{1/2}$$

$$N_T = \sqrt{N_R^2 + N_{time}^2} = \left(N_R^2 + \tau(FA\epsilon Q_e + i_{DC} + F_\beta A\epsilon Q_e \Omega)\right)^{1/2} = \left(N_R^2 + \tau n_t^2\right)^{1/2}$$

We define the telescope effective area $A_\epsilon$

$A_\epsilon = A\epsilon Q_e$ note that this includes the quantum efficiency

$$\frac{S}{N} = \frac{S}{N_T} = \frac{FA_\epsilon \tau}{\left[N_R^2 + \tau(FA_\epsilon + i_{DC} + F_\beta A_\epsilon \Omega)\right]^{1/2}} = \frac{FA_\epsilon \sqrt{\tau}}{\left[\frac{N_R^2}{\tau} + n_t^2\right]^{1/2}} = \frac{FA_\epsilon \tau}{\left[N_R^2 + \tau n_t^2\right]^{1/2}}.$$

At short time scales the S/N increases linearly with integration time $\tau$ since the read noise dominates the noise and then transitions to increasing as $\tau^{1/2}$ at increasing times as the shot noise from the source backgrounds and dark current begin to dominate. We define the transition time between these two domains as $\tau_c$.

$$\tau_c = \frac{N_R^2}{FA_\epsilon + i_{DC} + F_\beta A_\epsilon \Omega} = \frac{N_R^2}{n_t^2}$$

$$S/N(\tau = \tau_c) = \frac{FA_\epsilon N_R}{\sqrt{2}(FA_\epsilon + i_{DC} + F_\beta A_\epsilon \Omega)} = \frac{FA_\epsilon N_R}{\sqrt{2} n_t^2}$$

We solve for the time to achieve a given S/N as follows:

$$S_N \equiv S/N = FA_\epsilon \tau / \left[N_R^2 + \tau n_t^2\right]^{1/2}$$

$$S_N^2 (N_R^2 + \tau n_t^2) = F^2 A_\epsilon^2 \tau^2$$

$$F^2 A_\epsilon^2 \tau^2 - S_N^2 n_t^2 \tau - S_N^2 N_R^2 = 0$$

$$\tau = \frac{S_N^2 n_t^2 \pm \sqrt{S_N^4 n_t^4 + 4F^2 A_\epsilon^2 S_N^2 N_R^2}}{2F^2 A_\epsilon^2} = \frac{S_N^2 n_t^2}{2F^2 A_\epsilon^2}\left[1 + \sqrt{1 + \frac{4F^2 A_\epsilon^2 N_R^2}{S_N^2 n_t^4}}\right]$$

We note that the computation of the SNR above assumes the shot noise from the source also contributes to the noise term. This is reasonable IF the SNR is computed relative to the pixel the source is detected in but in most search strategies we will be doing spatial filtering and the SNR should be computed relative to the nearby pixels that do not have the source term in them. Thus in the above we set $n_t = \left(i_{DC} + F_\beta A_\epsilon Q_e \Omega\right)^{1/2}$ (noise in pixels outside source) in computing the SNR, $\tau$ and $\tau_c$ for comparing SNR of a possible source to its nearby pixels without the source. This increases the effective SNR so that $\frac{S}{N} = \frac{FA_\epsilon \tau}{\left[N_R^2 + \tau n_t^2\right]^{1/2}} = \frac{FA_\epsilon \tau}{\left[N_R^2 + \tau\left(i_{DC} + F_\beta A_\epsilon \Omega\right)\right]^{1/2}}$.

## 7. SIMULATIONS

We compute some examples of the SNR for differing civilization classes, distances and hence redshifts for existing or soon to exist telescopes below. He we include all backgrounds and modest seeing and detectors but assume we have narrow enough filters to exclude airglow and OH lines or that we are in regions where these are minimal.

A more aggressive approach is to assume we use the same technology to receive as is used to transmit by the civilization target. In the latter case the SNR becomes extremely large across the entire horizon for space based surveys using this approach with civilizations of class 3 and above assuming we also become a class 3 civilization.

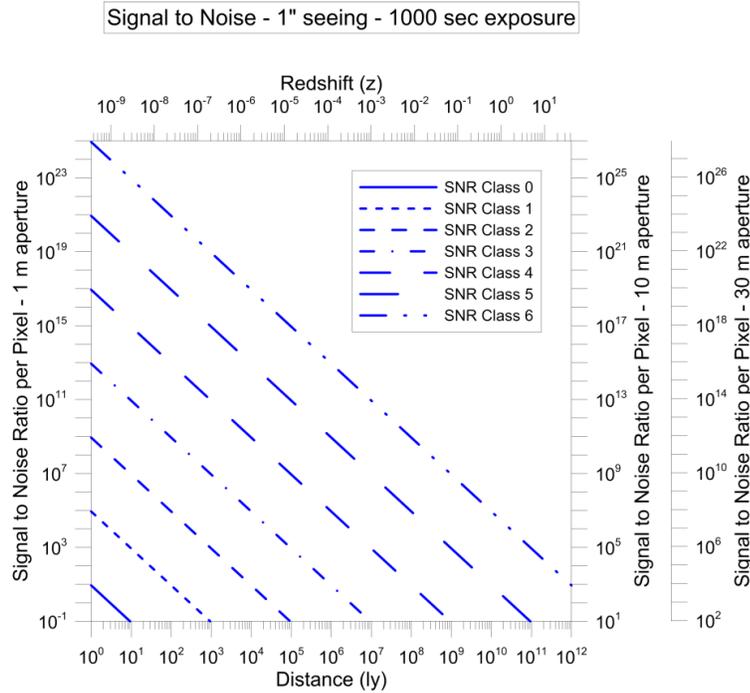

**Figure 19 -** SNR vs distance and civilization class for a 1, 10, 30 meter ground based telescope with a 1000 sec integration and very modest system assumptions with seeing of 0.5" RMS without adaptive optics, pixel size of 0.5", readout noise of 10e, dark current of 1e/s, QE =0.5 and atmospheric transmission of 0.5. The total noise is dominated by the readout noise and dark current and relatively insensitive to bandwidth. This represents our current (or soon to exist) capability for modestly wide field imaging. Our current technology for adaptive optics would be useful for narrow spatial surveys or follow up but is not currently feasible for wide field (degree class) surveys. The bottom line for even 1 m class telescopes is that class 3 civilization are detectable across our galaxy, class 4 civilizations are detectable in nearby galaxies and class 5 civilization are detectable out to modest redshifts. With 10 and 30 m class telescopes the situation is even more optimistic. A wide field LSST like telescope (8m class) could detect class 4 civilizations out to high redshift. We can reduce the readout noise to 1e and the dark current to negligible levels if needed and can enter a photon counting regime for narrow bandwidth cases.

### 7.1 – Space based options

While ground based options are the least costly to implement, space based approaches offer a number of advantages. There is no atmospheric windows to deal with and the backgrounds drop to the zodiacal light limit (assuming missions within the solar system). This can dramatically open up the wavelength search space and offer much greater sensitivity for a given aperture. The main issue is cost, complexity and aperture limit. The same system we propose for the phased array transmission in DE-STAR can be used in a bidirectional mode as a receiver as well. If we imagine we expand our space based capability so that we become a class 2,3 or 4 civilization the ability to detect other distant civilization becomes much greater. One disadvantage in the phased array receive mode is that the simplest designs are single pixel and thus we lose the advantage of spatial multiplexing. There are future approaches to the spatial multiplexing problem but they are not yet practical. Ground based variants of this approach also offer the possibility of extremely large aperture though with limited number of spatial pixels.

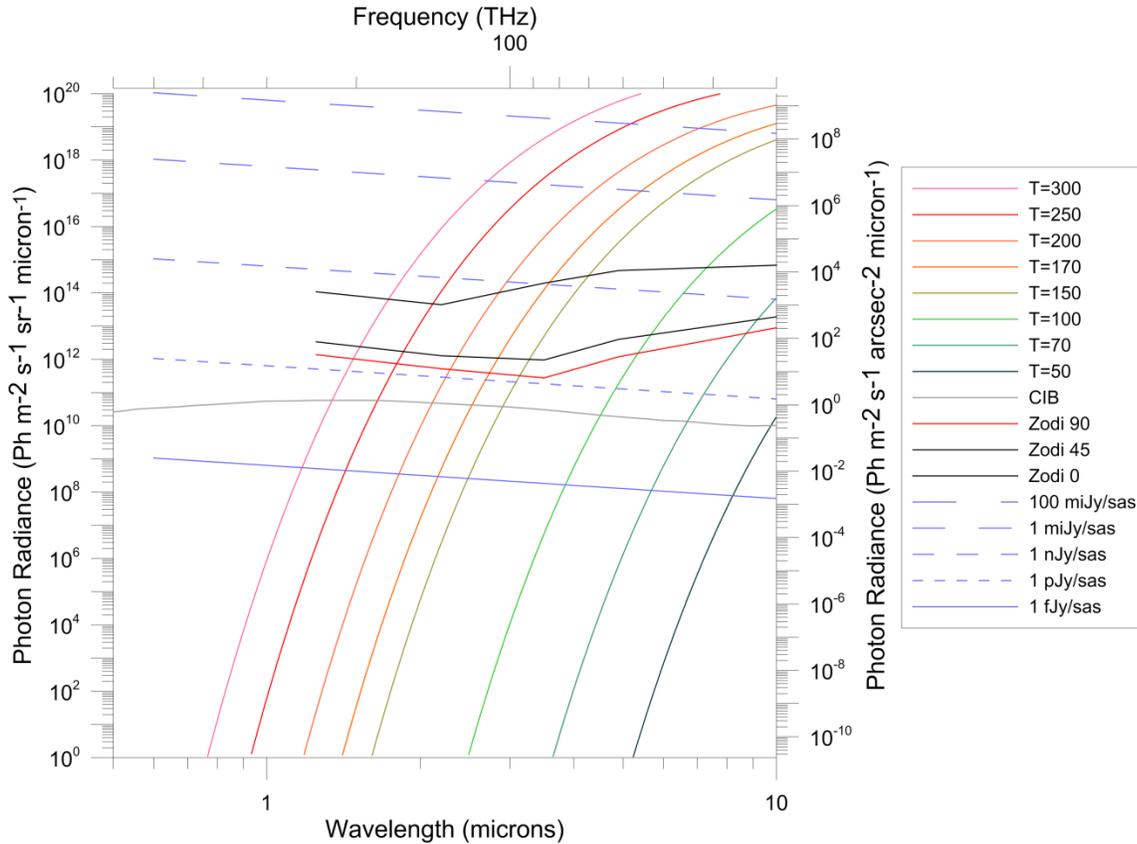

**Figure 20 -** Space based mission thermal emission from optics, CIB and Zodiacal light in the ecliptic plane (0) at 45 degrees relative to the plane (45) and perpendicular to the plane (90). Zodi is for COBE DIRBE day 100. The term "sas" refer to square arc second.

### 7.2 – Effects of filter bandwidth on SNR – filter optimization

The filter bandwidth affects the SNR since the wider the filter the greater the background light accepted. The wider the filter the less bands are needs to cover a broad range of possible laser lines but the less wavelength specificity and the poorer the SED modeling possible. If we focus on the SNR while parameterizing the various backgrounds and detector noise terms we can compute the effect of varying the filter bandwidth.
We write the noise contribution as above:

$$N_T = \sqrt{N_R^2 + N_{time}^2} = \left(N_R^2 + \tau(FA_\epsilon + i_{DC} + F_\beta A_\epsilon \Omega)\right)^{1/2}$$
$$\rightarrow \left(N_R^2 + \tau(i_{DC} + F_\beta A_\epsilon \Omega)\right)^{1/2}$$

where we have removed the photon statistics of the source itself since we are comparing to nearby pixels without the source noise.

$$\frac{S}{N} = \frac{FA_\epsilon \tau}{\left[N_R^2 + \tau(i_{DC} + F_\beta A_\epsilon \Omega)\right]^{1/2}}$$

For narrow bandwidths $\Delta\lambda$, we can write the photon flux terms F ($\gamma$/s-m$^2$) and $F_B$ ($\gamma$/s-m$^2$-st) as:
F = $\underline{B}\Delta\lambda$ and $F_B = \underline{B_B}\Delta\lambda$ where $\underline{B}$($\gamma$/s-m$^2$-$\mu$) and $\underline{B_B}$ ($\gamma$/s-m$^2$–st-$\mu$) are per unit bandwidth.

This gives a total noise term (in pixels away from signal) of $N_T = \left(N_R^2 + \tau(i_{DC} + B_\beta \Delta\lambda A_\epsilon \Omega)\right)^{1/2}$

In this case the S/N between the signal pixels and the non signal pixels is:

$$\frac{S}{N} = \frac{FA_\epsilon \tau}{N_T} = \frac{FA_\epsilon \tau}{\left[N_R^2 + \tau(i_{DC} + B_\beta \Delta\lambda A_\epsilon \Omega)\right]^{1/2}}$$

From this we see that for very small filter bandwidths the background contribution to the noise term $\tau B_\beta \Delta\lambda A_\epsilon Q_e \Omega$ is negligible and it is only at long integration times with small dark currents and large backgrounds $B_\beta$ that this bandwidth dependent term becomes important. Note that $B_\beta$ includes all sources of background except the detector. These include the telescope emission, atmosphere including air glow and OH lines, Zodiacal light, unresolved stars and the CIB. The background can become large due to OH emission as well as optical and atmospheric thermal emission in the IR, especially beyond 2.4 microns. This is where very narrow bandwidth filter will be very helpful even though OH lines will remain until the filter bandwidth becomes extremely narrow (essentially an IFU) where we can then observe between the OH lines. Beyond 2.5 microns there is little OH emission as discussed previously. See the discussion and plots above.

When we reach the level of a total noise, in our integration time, of roughly 1 electron there is little reason to go lower. Since we rapidly become signal photon starved, for modest civilization classes at large distances, there is a premium on low readout noise devices to achieve one electron of (including detector) noise. With modern detector arrays and narrow band filters it is feasible to approach this level of noise. When observing nearby bright galaxies in the core regions with the most stars the effective background due to the unresolved (but bright) star light can be a significant background term and here reducing the filter bandwidth is important. It is the relative relationship between the read noise, the dark current and the background term that is critical to understand to optimize the filter. If telescope time is not an issue and if filter costs are not important then a very narrow bandwidth filter is preferable.

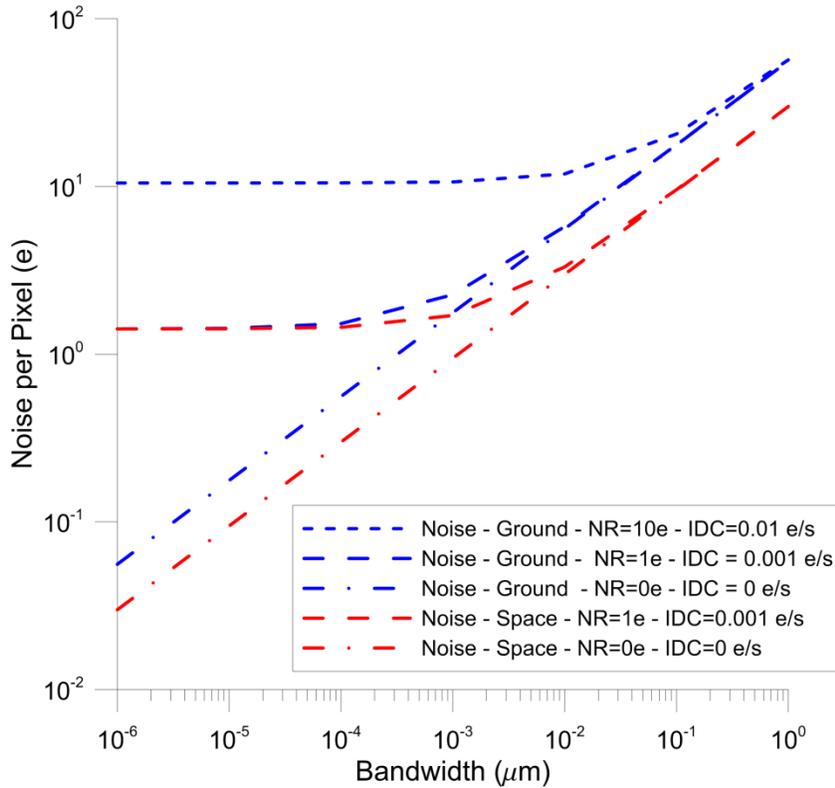

**Figure 21** – Noise per pixel vs filter bandwidth for several hypothetical wide field cases. Includes detector noise is given as well as background noise. For small filter bandwidths the noise is detector noise limited while for large bandwidth the noise is background limited. At very large bandwidth the noise converges and is proportional to $\beta^{1/2}$ where $\beta$ is the filter bandwidth. For a space mission the measurements are limited by zodiacal light and detector noise, assuming the optics are cooled sufficiently (for the IR cases) as to not dominate. In general a space mission will used diffraction limited optics and there will be no atmospheric "seeing issues". For the ground based cases we assume a wide field system and we assume adaptive optics cannot be used over the wide FOV. We also show a noiseless detector case for reference which could be realized for photons counting systems such as superconducting MKIDs (Microwave Kinetic Inductor Detector) or possible advanced cooled APD (Avalanche Photo Diode) arrays. Neither is currently available in the large formats ideally needed. For the ground case we assume the atmospheric transmission is 0.5, the pixel size is 0.5", the seeing is 1", the total telescope optical efficiency is 0.5, the detector QE = 0.5 and the atmospheric and extraterrestrial background is 100 $\gamma$/s-m$^2$-µm-sq-arcsec. Note that depending on the wavelength and the sky conditions the background could be significantly larger especially in the presence of OH lines in systems with low resolving power (wider filter bandwidth). For the space based case we assume the system is diffraction limited, the total telescope optical efficiency is 0.7, the detector QE = 0.8 and the extraterrestrial background is 10 $\gamma$/s-m$^2$-µm-sq-arcsec and that the optics are sufficiently cooled. The primary advantage of space is the lack of atmospheric emission, particularly of OH lines in J and H bands as well as the ability to cool the optics for K band and beyond.

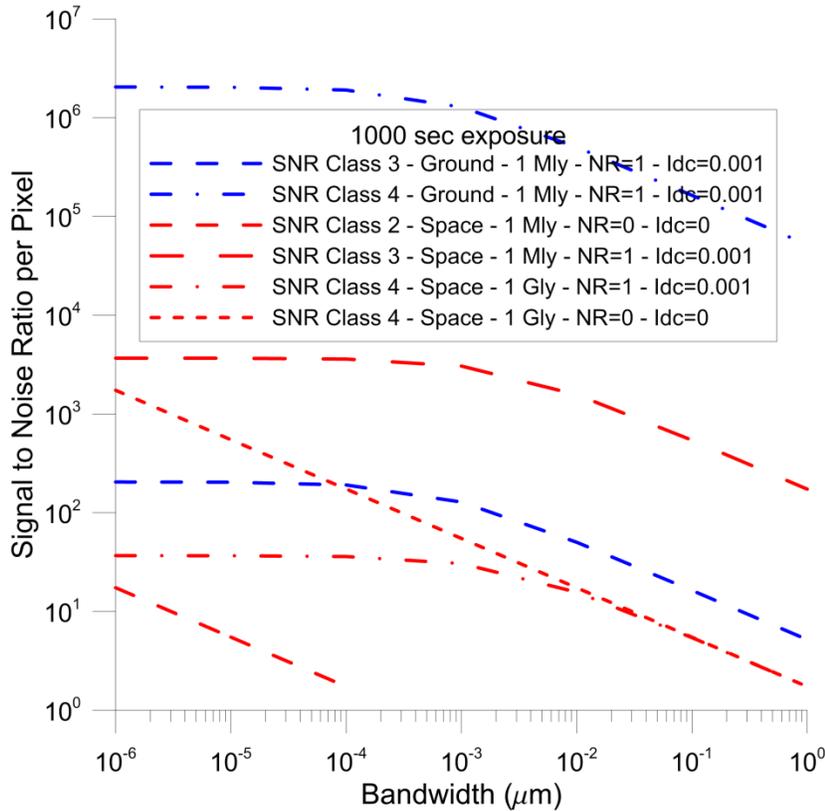

**Figure 22** – Signal to noise ratio for several ground and space based scenarios with a 1 meter aperture at the Earth. NR is the detector readout noise in e and Idc is the detector dark current in e/s. For the ground case we assume the atmospheric transmission is 0.5, the pixel size is 0.5", the seeing is 1", the total telescope optical efficiency is 0.5, the detector QE = 0.5 and the atmospheric and extraterrestrial background is 100 γ/s-m$^2$-μm-sq-arcsec. Note that depending on the wavelength and the sky conditions the background could be significantly larger especially in the presence of OH lines in systems with low resolving power (wider filter bandwidth). For the space based case we assume the system is diffraction limited, the total telescope optical efficiency is 0.7, the detector QE = 0.8 and the extraterrestrial background is 10 γ/s-m$^2$-μm-sq-arcsec and that the optics are sufficiently cooled.

### 7.3 – Effects of Pixel size on SNR

In analogy with the discussion above of the effects of the filter bandwidth on the noise and SNR we now apply the same formalism to the effects of the pixel size. The pixel size affects the noise and SNR since the wider the pixel size the greater the background light accepted. The pixel size ($\theta$) is the full angle of a pixel and the solid angle of the pixel is related simply as (for small angles) $\Omega = \theta^2$ We write the noise contribution and SNR as above:

$$N_T = \sqrt{N_R^2 + N_{time}^2} = \left(N_R^2 + \tau(FA\epsilon Q_e + i_{DC} + F_\beta A\epsilon Q_e \Omega)\right)^{1/2}$$
$$\rightarrow \left(N_R^2 + \tau(i_{DC} + F_\beta A\epsilon Q_e \Omega)\right)^{1/2}$$

where we have removed the photon statistics of the source itself since we are comparing to nearby pixels without the source noise.

$$\frac{S}{N} = \frac{FA_\epsilon \tau}{\left[N_R^2 + \tau(i_{DC} + F_\beta A_\epsilon \Omega)\right]^{1/2}}$$

We use the same notation where we write the photon flux terms F ($\gamma$/s-m$^2$) and F$_B$ ($\gamma$/s-m$^2$-st) as: F = $\underline{B}\Delta\lambda$ and F$_B$ = $\underline{B_B}\Delta\lambda$ where $\underline{B}$($\gamma$/s-m$^2$-µ) and $\underline{B_B}$ ($\gamma$/s-m$^2$–st-µ) are per unit bandwidth.

This gives a total noise term (in pixels away from signal) of $N_T = \left(N_R^2 + \tau(i_{DC} + B_\beta \Delta\lambda A\epsilon Q_e \Omega)\right)^{1/2}$

In this case the S/N between the signal pixels and the non signal pixels is:

$$\frac{S}{N} = \frac{FA_\epsilon \tau}{N_T} = \frac{FA_\epsilon \tau}{\left[N_R^2 + \tau(i_{DC} + B_\beta \Delta\lambda A\epsilon Q_e \Omega)\right]^{1/2}}$$

In analogy with spectral bandwidth we see that for very small pixel sizes the background contribution to the noise term $\tau B_\beta \Delta\lambda A\epsilon Q_e \Omega$ is usually negligible and it is only for very large pixels at long integration times with small dark currents and large backgrounds $B_\beta$ that this solid angle dependent term becomes important. As before $B_\beta$ includes all sources of background except the detector. These include the telescope emission, atmosphere thermal and lines (air glow) and OH lines (if inside the atmosphere based), Zodiacal light, unresolved stars and the CIB. In the near IR the background can become large due to OH emission as well as optical and atmospheric thermal emission in the IR, especially beyond 2.4 microns. Beyond 2.5 microns there is little OH emission as discussed previously.

The obvious question is why would we want large pixels? The answer is the following. In some search scenarios we are looking for any source of anomalous spectral emission and IF we use a high resolving power spectrometer with a wide "pixel" we might be able to leverage the spectral resolution to get to lower backgrounds by observing between the "lines" and cover a larger field of view (large pixel) and hence multiplex the observation by looking at a larger number of sources. This trades off spatial resolution for spectral resolution but with the ability to use a spectrometer. This would be an unusual spectrometer which present challenges in construction but may allow a higher thruput in some circumstances. Functionally this could be a larger fiber spectrometer.

Note that the background contribution to the noise term $\tau B_\beta \Delta\lambda A\epsilon Q_e \Omega$ is proportional to the product of spectral bandwidth, aperture area and pixel solid angle $\Delta\lambda\Omega = \Delta\lambda\theta^2$ and hence we have the same scaling of noise and SNR with bandwidth and with solid angle.

In the case of a diffraction limited telescope we note the background contribution to the noise term $\tau B_\beta \Delta\lambda A\epsilon Q_e \Omega$ is proportional to $\Delta\lambda A\Omega$. For a diffraction limited telescope the diffraction limited pixel size (not over sampled) is such that $A\Omega = \lambda^2$ and hence the background noise contribution term is $\tau B_\beta \Delta\lambda A\epsilon Q_e \Omega = \tau B_\beta \Delta\lambda\epsilon Q_e \lambda^2$.

OH lines are unresolved at R=10,000 so there will be a practical tradeoff between observing between OH and air glow lines and fractional "clean" spectral coverage. R=1000 to 10,000 is generally a practical range. Spectral cross talk with larger pixels will likely be an issue to be explored. Ideally both a spectrometer and an spatial search using narrow band filters would be employed. This is allow both rapid and deep searches as well as integral follow up and filtering of atmospheric lines (for ground based systems).

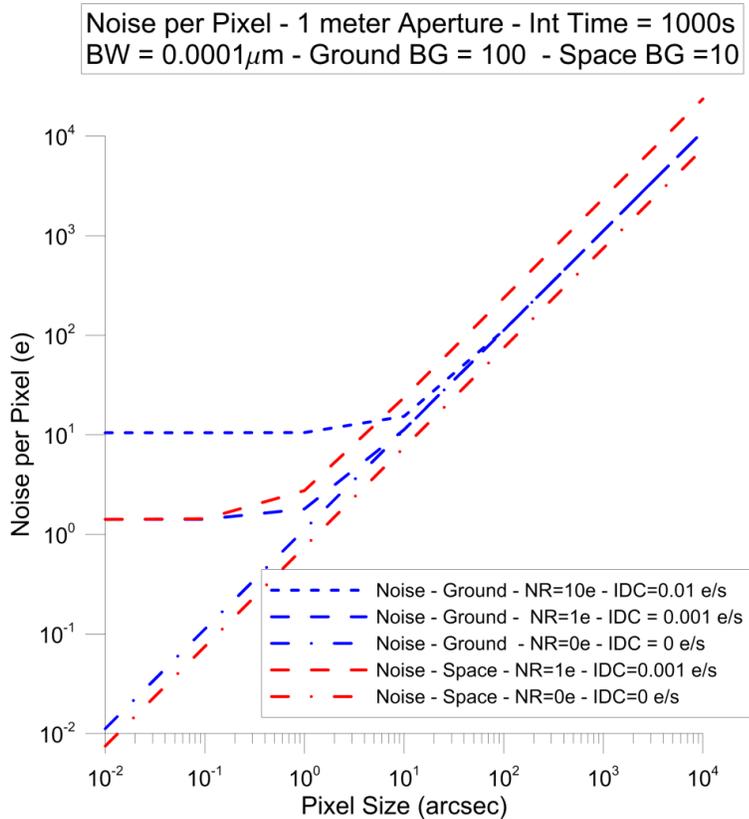

**Figure 23 -** Noise per spectrometer pixel vs pixel size for several hypothetical cases where a high resolution spectrometer is used (1 Angstrom – R=10,000 at 1 micron). We assume observations are made near 1 micron (visible to near IR). Includes detector noise is given as well as background noise. For a small spectrometer "pixel" the noise is detector noise limited while for large "pixel" input the noise is background limited. At very large pixel size the noise converges and is proportional to the pixel size. For a space mission the measurements are limited by zodiacal light and detector noise, assuming the optics are cooled sufficiently (for the IR cases) as to not dominate. In general a space mission will use diffraction limited optics and there will be no atmospheric "seeing issues". For the ground based cases we assume a seeing of 1 arc sec. The small pixel values for the ground case (smaller than seeing) are not relevant. We also show a noiseless detector case for reference which could be realized for photons counting systems such as superconducting MKIDs or possible advanced cooled APD arrays. Neither is currently available in the large formats ideally needed. For the ground case we assume the atmospheric transmission is 0.5, the total telescope optical efficiency is 0.5, the detector QE = 0.5 and the atmospheric and extraterrestrial background is 100 $\gamma$/s-m$^2$-µm-sq-arcsec. Using a high resolving power spectrometer allows us to observe with low background between the atmospheric telluric lines. Note that depending on the wavelength and the sky conditions the background could be significantly larger especially in the presence of OH lines.. For the space based case we assume the system is diffraction limited, the total telescope optical efficiency is 0.7, the detector QE = 0.8 and the extraterrestrial background is 10 $\gamma$/s-m$^2$-µm-sq-arcsec and that the optics are sufficiently cooled.

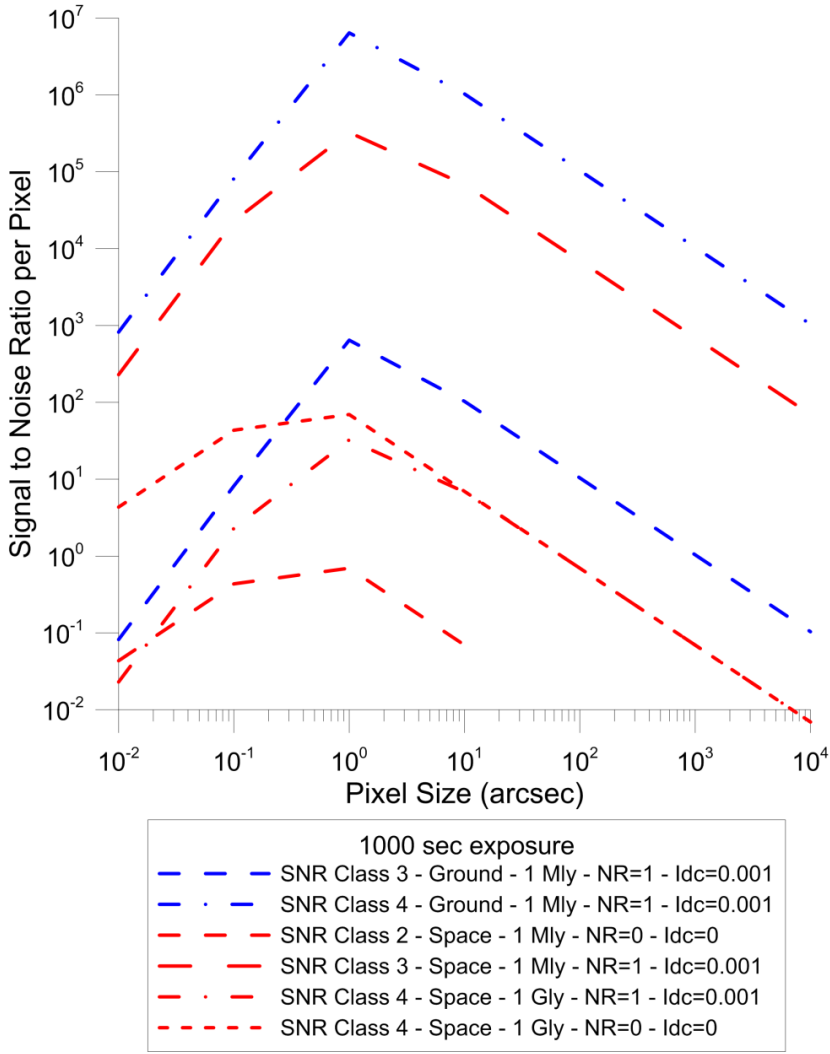

Figure 24 - Signal to noise ratio vs pixel size for a high resolution spectrometer (0.1nm , R=10,000 at 1 micron) for several ground and space based scenarios with a 1 meter aperture. Larger pixels are background limited and smaller ones are detector limited. NR is the detector readout noise in e and Idc is the detector dark current in e/s. For the ground case we assume the atmospheric transmission is 0.5, the pixel size is 0.5", the seeing is 1", the total telescope optical efficiency is 0.5, the detector QE = 0.5 and the atmospheric and extraterrestrial background is 100 $\gamma$/s-m$^2$-µm-sq-arcsec. Note that depending on the wavelength and the sky conditions the background could be significantly larger especially in the presence of OH lines in systems with low resolving power (wider filter bandwidth). For the space based case we assume the system is diffraction limited, the total telescope optical efficiency is 0.7, the detector QE = 0.8 and the extraterrestrial background is 10 $\gamma$/s-m$^2$-µm-sq-arcsec and that the optics are sufficiently cooled. The peak in the SNR is an due to the diffraction limit for a 1 meter aperture at 1 micron and thus the signal is spread out over pixels smaller than the PSF. The peak of the SNR is indicative of an optimally matched spectrometer FOV. A larger aperture with adaptive optics on the ground or a larger space based telescope would peak at smaller pixel sizes. This plot is indicative of the performance of a single pixel high R spectrometer. An IFU could also be used to more optimally both spatially and spectrally search.

We have a choice of how long we will integrate an image for before reading out. This is the integration time τ. As long as the integration time is longer than the time needed to get the required SNR then it is sufficient. For shorter

**7.4 Near term future facilities**

In addition to the existing ground and space based assets we will soon have wide field ground based capability with the LSST in the visible and near IR and excellent, though very narrow FOV space based capability with JWST out to 28 microns. In addition we will have the ground based 30 m class telescopes again with narrow FOV. All of these will be available in the next decade if all goes as planned. All of the above analysis applies to the ground based LSST and 30m class telescopes and with the expanding IFU and related spectroscopy this gives excellent follow up capability to possible detection with wide field instruments like the LSST among others. The observation strategy for an effective SETI search would need to be modified for optimum use in the case discussed here.

JWST allows for a qualitatively new capability as the wavelength range is greatly extended compared to ground based assets. With spectroscopic capability this allows for unique opportunities though the narrow FOV is a problem for blind search strategies. JWST also offers the possibility of greater redshift space coverage for a given (though unknown) transmit wavelength.

**7.5 – Civilizations with comparable transmit and receive capabilities**

As mentioned our civilization currently the equivalent of about 1.5. Rapid progress to civilization class 4 is feasible within 50 years if the will existed to do so. Since the basic technology we propose is bidirectional and can operate in both a transmit and receive mode, we now ask what the quantitative consequences of this are. We apply the same methodology as above for existing small ground and space based telescopes but focus on space based deployment. The bottom line is that detection across the entire horizon is feasible with the usual caveat of being in the relevant band for detection.

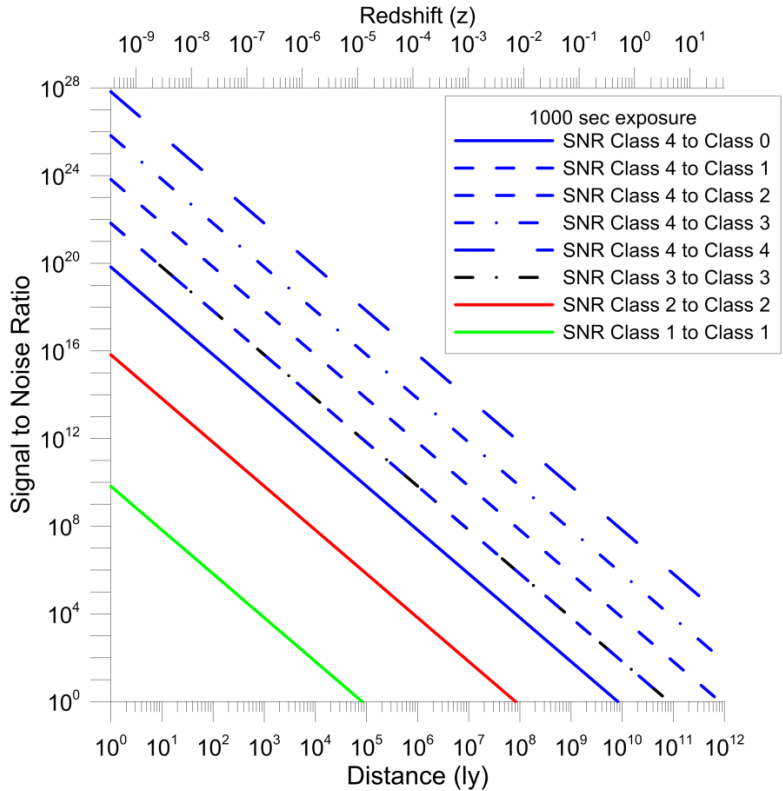

**Figure 25** – SNR for diffraction limited space based arrays of various civilization classes vs distance for a single 1000 sec integration. The first class is the transmitter and the second is the receiver (Earth). Space background is assumed to be 10 $\gamma$/s-m$^2$-µm-sq-arcsec. Telescopes are assumed to be ideal and the detector is assumed to be photon counting. The receive bandwidth is 1nm wide. $10^{12}$ ly (~ 300 Gpc) corresponds to a redshift z~20.

## 7.6 – Blind searches and blind transmission – optimizing strategies

A major question in all searches is "why would "they" transmit towards us"? The equivalent for us is "why would we "look" at them"? In the case of "both sides" within our galaxy we already have preferred directions towards known exoplanets, though these appear to be ubiquitous through our own galaxy and presumably others. "We" could look towards known higher probability candidates based on presumed habitability for life and "they" could do the same. Since we are on the very beginnings of searching for exoplanets we can imagine a more advanced civilization would have vastly more knowledge of likely targets to transmit to. As we go beyond our own local realm and begin looking at extragalactic targets "we" could look towards all nearby galaxies. "They" could do the same. As we go to high redshift targets we have little to guide us at our current level of knowledge. We could look towards galaxies with age distributions we deem more probable for the formation of life as one example. In our case using the DE-STAR phased array as the transmitter we can send out multiple beams or time share between beams to optimize chances for detection. This is a probability "game" for which we do not know the real "rules of the game" explicitly. The reality is we have little quantitative information to use so we enter the realm of blind searches.

The flux at the Earth from a civilization class S at distance L is $F (\gamma/s\text{-}m^2) = \xi P/(L \theta)^2 = \xi P/L^2 \Omega$ where $\theta$ and $\Omega$ are the transmitted beam divergence angle and solid angle respectively and $\xi = (hc/\lambda)^{-1}$. Here $P(w) = F_E \epsilon_c 10^{2S}$ and $\theta = 2 \lambda/d$ where $d(m) = 10^S$ with $\theta = 2 \lambda 10^{-S}$ and $\Omega = \theta^2 = (2\lambda/d)^2 = 4 \lambda^2 10^{-2S}$ where $F_E$ is the solar insolation at the top of the Earth's atmosphere or $F_E \sim 1400$ w/m$^2$.

$F (\gamma/s\text{-}m^2) = \xi P/L^2 \Omega = \xi F_E \epsilon_c 10^{2S}/(L^2 4 \lambda^2 10^{-2S}) = \xi F_E \epsilon_c 10^{4S}/(4L^2\lambda^2) = (hc)^{-1} F_E \epsilon_c 10^{4S}/(4L^2\lambda)$. We can immediately see why going to shorter wavelengths (for constant transmission power and array size) increases the photon flux (two powers of $\lambda$ from diffraction and one power (inverse) from photon energy. The received flux in w/m$^2$ is $F_w (w/m^2) = F_E \epsilon_c 10^{4S}/(4L^2\lambda^2)$.

The forward gain of antenna is $G = 4\pi/\Omega = \pi d^2/\lambda^2 = 4\pi/(4\lambda^2 10^{-2S}) = \pi \lambda^{-2} 10^{2S}$. The antenna gain is equal to the number of "non-overlapping" beams on the sphere. The gain for a class 4 system operating at 1μ is over 200 db. This is useful in comparing microwave to optical/IR SETI where we see the forward gain of the optical system is vastly greater (by the ratio of wavelengths squared) than the equivalent sized microwave system. One could argue that microwave systems are much easier to build in larger apertures than optical ones to counter this and indeed we have very large microwave telescopes (100m class) but only 10m class optical telescopes, however this still does not give the gain that an optical/ IR telescope has.

If we assume the transmission comes from a phased array (our baseline) then the power can be distributed into (up to) as many beams as there are array elements. If we assume the transmitted beam is split into N beams (one of which is incident on the Earth) then we have ($\Omega_N = N \Omega$ is the split beam solid angle) and the new received flux $F_N$ is:

$F_N (\gamma/s\text{-}m^2) = \xi P/N/(L^2 \Omega_N) = \xi P/(N^2 L^2 \Omega) = F/N^2 = \xi F_E \epsilon_c 10^{4S}/(4L^2\lambda^2)/N^2$. This obviously reduces the received flux but increases the transmitted solid angle $\Omega_N$ as there are N of these beams. This increases the probability of a "blind transmission" reception in terms of number of beam but at reduced flux. Depending on the type of search strategy on the received side these can essentially cancel out. This depends on the time gating of the reception strategy. Time multiplexing on the transmit side (beam switching not data encoding) is another strategy for transmission. A phased array is ideal for rapid beam switching. The same phased array transmission system is also a phased array receiver but we do not assume this at our current level of detection strategy.

**Mapping speed** – Another way of thinking about blind search strategies is to look at the SNR (Fig 25) for a given civilization class $S_t$ that is transmitting and a civilization class that is receiving $S_r$. Recall Figure 25 is for a single 1000s observation. In the above discussion the civilization class "S" is the transmitting class $S_t$. The mapping speed (ie how many sources or how much sky area (civilizations of class $S_t$ that can be mapped to a given SNR in integration time τ by receiving civilization class $S_r$). Normally we are severely limited by the receiving telescope instantaneous FOV, but this is a technological limitation that could be overcome if we could develop optical phased arrays with optical correlators similar to that done in the radio (eg SKA). While this technology is not mature in the optical it could be developed and this would vastly increase our capability. This is similar to the transmit phased array but in reverse. This discussion is for another paper but should be considered. We note that the mapping speed is proportional to the square of the SNR in general **for a background limited** detection or to just linear in the SNR in the non background limited case. As discussed above there is a transition from the linear increase in SNR with time at short time scales when we are NOT background limited but read noise limited to the slower $\tau^{1/2}$ increase with time when we are dark current and background limited regime. The critical transition time scale $\tau_c$ from linear increase to $\tau^{1/2}$ we defined was given by (for the case of comparing the signal pixels to the surrounding "noise pixels":

$$\tau_c = \frac{N_R^2}{i_{DC} + F_\beta A_\epsilon \Omega} = \frac{N_R^2}{n_t^2}$$

$$S/N(\tau = \tau_c) = \frac{FA_\epsilon N_R}{\sqrt{2}(i_{DC} + F_\beta A_\epsilon \Omega)} = \frac{FA_\epsilon N_R}{\sqrt{2}n_t^2}$$

$$n_t^2 = i_{DC} + F_\beta A_\epsilon \Omega$$

Whether we are in the linear (with time) SNR regime for non background limited detection or the $\tau^{1/2}$ regime where we are background limited needs to be determined to consider mapping speed scaling. For the case of extremely low readout out noise ($N_R$) and dark current ($i_{DC}$) (the case for superconducting detectors) we will be background limited and hence in the $\tau^{1/2}$ regime while for larger $N_R$ but still low dark current $i_{DC}$ we will be readout noise dominated and not background dominated, especially since we have extremely narrow receive bandwidths) and hence in the linear regime (SNR ~ $\tau$). In many cases where we have extremely large SNR's we are NOT background limited and hence SNR ~ $\tau$.

From Fig 25 we see extremely large SNR's are possible depending on the civilizations: $S_t$ and $S_r$. Figure 25 SNR implicitly assume the receiving civilization $S_r$ is "looking at" the transmitting civilization $S_t$ AND that the transmitting civilization is transmitting to the receiving civilization. In our case we are only ONE receiving civilization and hence the existential question we usually ask is "is anyone out there" – ie do we "see anyone" transmitting to us. A more general question is "what sees what" but this is of little general interest to us currently. For example, if we focus back on us and look at the SNR possible vs transmitting civilization class we see that a transmitting civilization class $S_t = 4$ and us - receiving class $S_r \sim 1.5$, we get possible SNR of $> 10^8$ for even large extragalactic distances ~ 100 Mpc (admittedly for space based observations – though ground in a narrow bandwidth can be close). Again this is for a single 1000 s observation. We need to ponder the time scale for total observations.

**Total observation times** – How long is a reasonable time to search for the existence of intelligence outside the Earth before we give up and admit we are alone? This is clearly a difficult question to ask and may require subsequent therapy. Some of the funding agencies answered this long ago. We will not discuss this part. However, it is not unreasonable to assume that the questions being asked are of great importance to some people (who need job security) and perhaps a human lifetime is not an unreasonable start. Let's just go for the time to tenure – say 35 years from BOL (beginning of life) to pick a simple number. This is approximately $10^9$ s or $10^6$ observations each of which is $10^3$ s. We can play this game a number of ways. Let us assume our current technology of a single mirror telescopes with limited FOV. Let's assume we can fill a telescope FOV with a single (but distant) galaxy. In our 30 year observing period we could point at $10^6$ galaxies (we will run out of galaxies at low z) but since we have an SNR of $>10^8$ for a single 1000s observation with the above assumptions ($S_t = 4$ receiving class $S_r \sim 1.5$) of an entire galaxy with some $10^{11}$ or more simultaneously observed planets) this seems somewhat wasteful. An SNR of $10^8$ allows us to multiplex in various ways with a multiplexing ratio of SNR (non background limited – typical for many of our cases) to $SNR^2$ (background limited). For the high SNR cases we are not background limited and hence the multiplex ratio is essential just the SNR. and linear in time τ. We can then imagine changing the integration time of detection or multiplexing the beams from the transmit side.

We do not need that many wavelength channels with perhaps $10^2$ -$10^4$ sufficing as discussed previously. We are still left with an extremely large multiplex factor left over. Here is where we can try to optimize wide and shallow vs narrow and deep observational strategies. for detection of different classes of civilizations.

We can calculate the probability that at a given redshift and given civilization class we will detect "them". We do this by assuming random pointing by the emitter (other civilizations) at their luminosity distance (redshift).

In the end we are brought to the conclusion, once again, that we already possess the ability to search for vast numbers of civilizations in modest time scales with modest telescopes. We do not have to be in space, though this would be more sensitive for a given aperture, and we do not even need 10m class telescopes so to make profound statements in this area. Referring back to our directed energy "Moore's Law" like evolution (Fig 1) we are immediately struck by the fact that for us, within a fraction of a human lifetime, we have already advanced enormously in terms of our ability to transmit (though we do not do so except as a by-product of our existence) and thus unless other advanced civilizations are "shy" like us they may also have greatly advanced in their ability to transmit. If we project but one human lifetimes into the future ($\sim 10^{-4 \text{to-} 5}$ of our human evolution) we see we will have the power to easily become a class 4 or greater civilization should we choose to do so and thus we will enter a period, as we are just beginning to now, where we can direct energy that is detectable by "others like ourselves" at essentially any reasonable redshift.

**Time to a desired SNR** – From the calculation above in section 7.6 we can compute the time τ required to a given SNR ($S_N$) as follows:

$$\tau_{SNR} = \frac{S_N^2 n_t^2 \pm \sqrt{S_N^4 n_t^4 + 4F^2 A_\epsilon^2 S_N^2 N_R^2}}{2F^2 A_\epsilon^2} = \frac{S_N^2 n_t^2}{2F^2 A_\epsilon^2}\left[1 + \sqrt{1 + \frac{4F^2 A_\epsilon^2 N_R^2}{S_N^2 n_t^4}}\right] = S_N^2 \frac{\tau_N}{2}\left[1 + \sqrt{1 + \frac{4}{S_N^2 \tau_N}\left(\frac{N_R}{n_t}\right)^2}\right]$$

$$= S_N^2 \frac{\tau_N}{2}\left[1 + \sqrt{1 + \frac{4}{S_N^2}\frac{\tau_c}{\tau_N}}\right]$$

where $\tau_N = \frac{n_t^2}{F^2 A_\epsilon^2}$, $\tau_c = \frac{N_R^2}{i_{DC} + F_\beta A_\epsilon \Omega} = \frac{N_R^2}{n_t^2}$ and $n_t = \left(i_{DC} + F_\beta A_\epsilon \Omega\right)^{1/2}$ units of $n_t$ in #$e^- - s^{1/2}$or #$e^- / Hz^{1/2}$

In the limit of no dark current and background noise $n_t = 0$ then $\tau_{SNR} = \dfrac{S_N^2 n_t^2}{2F^2 A_\epsilon^2}\left[1 + \sqrt{1 + \dfrac{4F^2 A_\epsilon^2 N_R^2}{S_N^2 n_t^4}}\right] = \dfrac{S_N N_R}{F\, A_\epsilon}$

In the limit of zero readout noise $N_R = 0$ then $\tau_{SNR} = \dfrac{S_N^2 n_t^2}{2F^2 A_\epsilon^2}\left[1 + \sqrt{1 + \dfrac{4F^2 A_\epsilon^2 N_R^2}{S_N^2 n_t^4}}\right] = \dfrac{S_N^2 n_t^2}{2F^2 A_\epsilon^2} = S_N^2 \dfrac{\tau_N}{2}$

Note the different scaling of $\tau$ in these two cases.

If $n_t = 0$ (read noise dominated case) then $\tau_{SNR} = \dfrac{S_N N_R}{F\, A_\epsilon} \sim \dfrac{S_N}{F}$

If $N_R = 0$ (background and/or dark current dominated case) then $\tau_{SNR} = S_N^2 \dfrac{\tau_N}{2} = \dfrac{S_N^2 n_t^2}{2F^2 A_\epsilon^2} \sim \dfrac{S_N^2}{2F^2}$

The flux at the Earth from a civilization class S at luminosity distance L is (from above) for one transmitted beam:

$F\ (\gamma/\text{s-m}^2) = \xi P/L^2 \Omega = \xi F_E \varepsilon_c\, 10^{2S}/(L^2\, 4\lambda^2 10^{-2S}) = \xi F_E \varepsilon_c\, 10^{4S}/(4L^2\lambda^2)$ with $\xi = (hc/\lambda)^{-1}$

With the beam solid angle $\Omega_{\text{beam}} = \theta^2 = (2\lambda/d)^2 = 4\lambda^2 10^{-2S}$ where $F_E$ is the solar insolation at the top of the Earth's atmosphere or $F_E \sim 1400$ w/m$^2$.

We can now compute the time required to achieve a given SNR for a given civilization class S (not to be confused with the SNR = $S_N$.

$\tau_{SNR} = S_N^2 \dfrac{\tau_N}{2}\left[1 + \sqrt{1 + \dfrac{4\,\tau_c}{S_N^2\, \tau_N}}\right]$ with $\tau_N = \dfrac{n_t^2}{F^2 A_\epsilon^2}$ and $\tau_c = \dfrac{N_R^2}{i_{DC} + F_\beta A_\epsilon \Omega}$

with the flux at the Earth from civilization class S being:

$F = \dfrac{1}{4}\xi F_E \varepsilon_c\, 10^{4S} L^{-2} \lambda^{-2}$

Thus $\tau_N = \dfrac{n_t^2}{F^2 A_\epsilon^2} = \dfrac{16 n_t^2 L^4 \lambda^4 10^{-8S}}{(\xi F_E \varepsilon_c)^2 A_\epsilon^2}$

Note the scaling $\tau_N \propto n_t^2, L^4, \lambda^4, 10^{-8S}, A_\epsilon^{-2}$

Thus $\tau_{SNR} = S_N^2 \dfrac{\tau_N}{2}\left[1 + \sqrt{1 + \dfrac{4\,\tau_c}{S_N^2\, \tau_N}}\right] = S_N^2 \dfrac{8 n_t^2 L^4 \lambda^4 10^{-8S}}{(\xi F_E \varepsilon_c)^2 A_\epsilon^2}\left[1 + \sqrt{1 + \dfrac{\tau_c}{S_N^2}\dfrac{(\xi F_E \varepsilon_c)^2 A_\epsilon^2}{4 n_t^2 L^4 \lambda^4 10^{-8S}}}\right]$

For a given civilization class that performs a blind beacon (transmission) we can compute the total solid angle of the sky they can cover in total time t for a given SNR (at the receiver – us) and luminosity distance L and thus the total number of possible planet targets. The total solid angle $\Omega_t$ the civilization could cover (in a beacon mode) in time t would be: $\Omega_t = (t/\tau)\,\Omega_{\text{beam}} = (t/\tau)\,4\lambda^2 10^{-2S}$ and hence the fraction of the sky covered by the transmitting civilization is $f_t = \Omega_t/4\pi$. This assumes the transmit beam is swept slowly enough that the "dwell time" on the target civilization is $\tau$. This is also complicated by the "transverse motion" of both the transmitting and receiving civilizations (see below). Note that this seems to decrease with larger S BUT the time $\tau$ to a given SNR is decreasing faster than the decrease in **beam solid angle** (since the power is rising) and hence the **total sky solid angle covered** in time "t" is **increasing with S as expected**. Of course, **the transmitting civilization has no idea what luminosity distance the receiving civilization is at, nor what reception capacity the receiving civilization has and thus "t" does not know about "τ"**. We can also interpret the transmit time "t" as being the total time of the Earth survey assuming the Earth based survey is a full sky "real time -full time" survey. If the Earth based survey is not full sky but covers a solid angle $\Omega_{rec}$ then the received "efficiency" is reduced by $f_r = \Omega_{rec}/4\pi$ and the effective Earth survey time required increases by $1/f_r = 4\pi/\Omega_{rec}$. An analogy to a lighthouse is appropriate here in the sense that the lighthouse notifies a distant ship of its presence. A problem is that is the lighthouse needs to be operational when the ship needs it AND the lighthouse must be transmitting. A complicating factor is the length of time the transmitting civilization is "on the air" which may be related to how long the civilization lasts. As is classic in SETI the number of "unknowns" greatly exceeds the number of "knowns". One strategy for the transmitting civilization is to transmit continuously to notify future civilizations of the existence of the (previous) transmitting civilization, since all receiving civilizations will receive in the future compared to the transmitting civilization. In this sense the "lighthouse" is left on indefinitely – "we will leave the lights on for you". The time of transmission or beacon "t" above will diverge and hence the solid angle covered will always be the full sky in this scenario. However, in order for the receiving civilization to receive, it must be either pointed at or enable a large or preferably a "real time – always on" full sky survey.

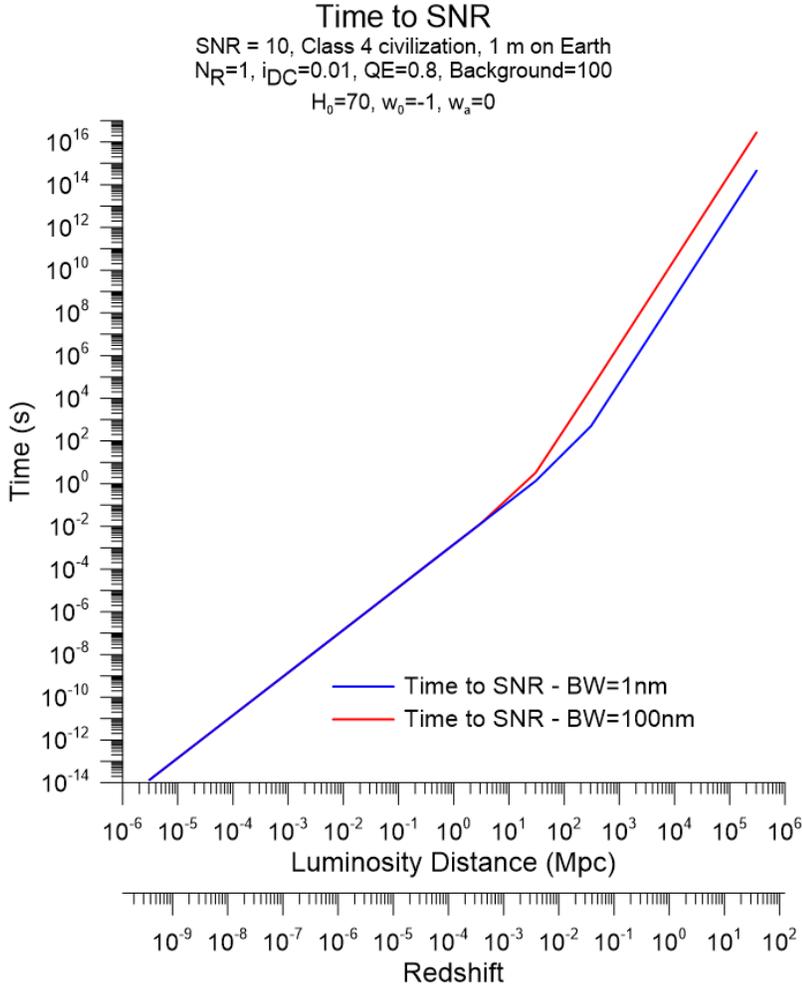

**Figure 26** – Time to SNR ($\tau_{snr}$) vs luminosity distance and redshift. Here the SNR is set to 10 for a 1m ground based wide field survey with an integration time of 1000 s and filter BW=1nm as well as BW=100nm. The transmitting civilization is Class 4. A benchmark or concordance model is used for the cosmological relationship between luminosity distance and redshift. Note that OH line emission is not included here as these depend on the specific wavelength.

**Effects of Integration Time on Noise and SNR -** We have a choice of how long we will integrate an image for before reading out. This is the integration time $\tau$. As long as the integration time is longer than the time needed to get the required SNR then it is sufficient IF the laser is in our beam for a time ($\tau_{laser}$) equal to or longer than $\tau$. For shorter integrations times we lose signal and hence SNR. If $\tau_{laser} > \tau$ we can integrate longer and since the signal is still "on" then we will increase our SNR. If $\tau_{laser} < \tau$ then signal is no longer "on" then increased integration time will only increase the noise and hence decrease the SNR. Since we do not know apriori what $\tau_{laser}$ is we want to understand the effects of changing $\tau$.

Recall the total noise in integration time $\tau$ in pixels nearby to the main source pixel is when the laser is on (in our beam) for time $\tau_{laser}$:

$$N_T = \sqrt{N_R^2 + N_{\text{time}}^2} = \left(N_R^2 + \tau(i_{DC} + F_\beta A\epsilon Q_e \Omega)\right)^{1/2} = \left(N_R^2 + \tau n_t^2\right)^{1/2}$$

where $n_t = (i_{DC} + F_\beta A\epsilon Q_e \Omega)^{1/2}$

And the SNR is:

$$SNR = \frac{S}{N} = \frac{S}{N_T} = \frac{FA_\epsilon \tau_{laser}}{\left[N_R^2 + \tau(i_{DC} + F_\beta A_\epsilon \Omega)\right]^{1/2}}$$

where $A_\epsilon = A\epsilon Q_e$

$$SNR = \frac{S}{N} = \frac{FA_\epsilon \tau_{laser}}{N_T} = \frac{FA_\epsilon \tau_{laser}}{\left[N_R^2 + \tau(i_{DC} + B_\beta \Delta\lambda A_\epsilon \Omega)\right]^{1/2}} = \frac{N_{e-laser}}{\left[N_R^2 + \tau(i_{DC} + B_\beta \Delta\lambda A_\epsilon \Omega)\right]^{1/2}}$$

where $F_\beta = B_\beta \Delta\lambda$ where $B_\beta (\gamma/s - m^2 - st - \mu) =$ background $\gamma$ per area, time, solid angle and wavelength

$N_{\gamma-laser} = FA_\epsilon \tau_{laser} =$ # electrons generated in "source" pixel during laser on time in beam of $\tau_{laser}$

Note that $N_{e-laser}$ can also include fast pulse cases as well as CW. Here there are now two times. We control the integration time τ but not the unknown "laser on in beam time" $\tau_{laser}$. We are trying to understand the effects of setting the integration time τ.

For a diffraction limited single mode (single polarization) system we have $A\Omega = \lambda^2$ and hence:

$$SNR_{DL} = \frac{S}{N} = \frac{FA_\epsilon \tau_{laser}}{N_T} = \frac{FA_\epsilon \tau_{laser}}{\left[N_R^2 + \tau(i_{DC} + B_\beta \Delta\lambda \epsilon Q_e \lambda^2)\right]^{1/2}} = \frac{N_{e-laser}}{\left[N_R^2 + \tau(i_{DC} + B_\beta \Delta\lambda \epsilon Q_e \lambda^2)\right]^{1/2}}$$

We clearly get the maximum SNR when $\tau = \tau_{laser}$ but since we do not know $\tau_{laser}$ we want to make τ larger than $\tau_{laser}$ but hence the difficulty. We do not know $\tau_{laser}$! The resolution is to understand how much the SNR degrades in various scenarios for different integration times. In addition we have a technical issue that most low noise devices have lower readout noise ($N_R$) if the readout time is not too short (ie pixel readout rate is slow enough to allow low noise electronics). Hence there is also a technical compromise depending on the device. As before when the integration time is short then we are readout noise dominated and when the integration time is long we are dark current and/ or background dominated. Modern low noise and low dark current devices can have $N_R \sim$ 1-10 e$^-$ and dark current $i_{dc}$ <0.01 e$^-$/s depending on operating temperature. If we use a photon counter like an APD (avalanche photo diode) then the detection is slightly different with dark rates being significantly higher but the time resolution being extremely short (typ ns). For a superconducting array detector like an MKID effectively $N_R = 0$ and $i_{dc} = 0$ though the array size is much smaller and the cryogenic requirements are much more complex.

We compute the total noise $N_T$ during integration time τ to help understand the effect of increased integration times. Here $N_T$ is given by:

$$N_T = \sqrt{N_R^2 + N_{\text{time}}^2} = \left(N_R^2 + \tau(i_{DC} + B_\beta \Delta\lambda A\epsilon Q_e \Omega)\right)^{1/2} = \left(N_R^2 + \tau n_t^2\right)^{1/2}$$

where $n_t = (i_{DC} + B_\beta \Delta\lambda A\epsilon Q_e \Omega)^{1/2}$

For the diffraction limited case $n_{t-DL} = (i_{DC} + B_\beta \Delta\lambda \epsilon Q_e \lambda^2)^{1/2}$

For a ground based telescope in the visible and near IR (if we avoid strong OH lines) we have $B_\beta \sim 100\ (\gamma/\text{s-m}^2\text{-arcsec}^2\text{-}\mu) \sim 4\times10^{12}(\gamma/\text{s-m}^2\text{-st-}\mu)$
For a space based asset we can have $B_\beta \sim 10\text{-}50\ (\gamma/\text{s-m}^2\text{-arcsec}^2\text{-}\mu) \sim 4\text{-}20\times10^{11}(\gamma/\text{s-m}^2\text{-st-}\mu)$

We can see the basic issues if we use a very low noise detector with $N_R = 1$ e⁻ and $i_{dc} = 0.01$ e⁻/s. If the bandpass is $\Delta\lambda = 0.01\mu m$ (10nm) then for the ground based case we have:
$N_T$ (e⁻) $\sim (1+\tau(0.01+1/\text{m}^2\text{-arcsec}^2))^{1/2}$. For a 1 m class telescope with 1 arc sec seeing we have:
$N_T \sim 1$ e⁻ for $\tau=1$ sec and $N_T \sim 30$ e⁻ for $\tau=1000$ sec

For the same system with $N_R = 10$ e⁻ and $i_{dc} = 0.01$ e⁻/s we have $N_T \sim 10$ e⁻ for $\tau=1$ sec and $N_T \sim 30$ e⁻ for $\tau=1000$ sec

Both of these are relatively small compared to the potential DE signal in many scenarios though clearly we would prefer a shorter integration time if $\tau_{laser}$ were short.
We discuss the spot dwell time for very distant beacons in terms of typical galactic motions. For a galactic survey we may prefer a shorter integration time of $\tau=1\text{-}10$ s while for higher redshift surveys we might prefer $\tau=1000$ s. We will see if makes little difference in detection probability in many cases.

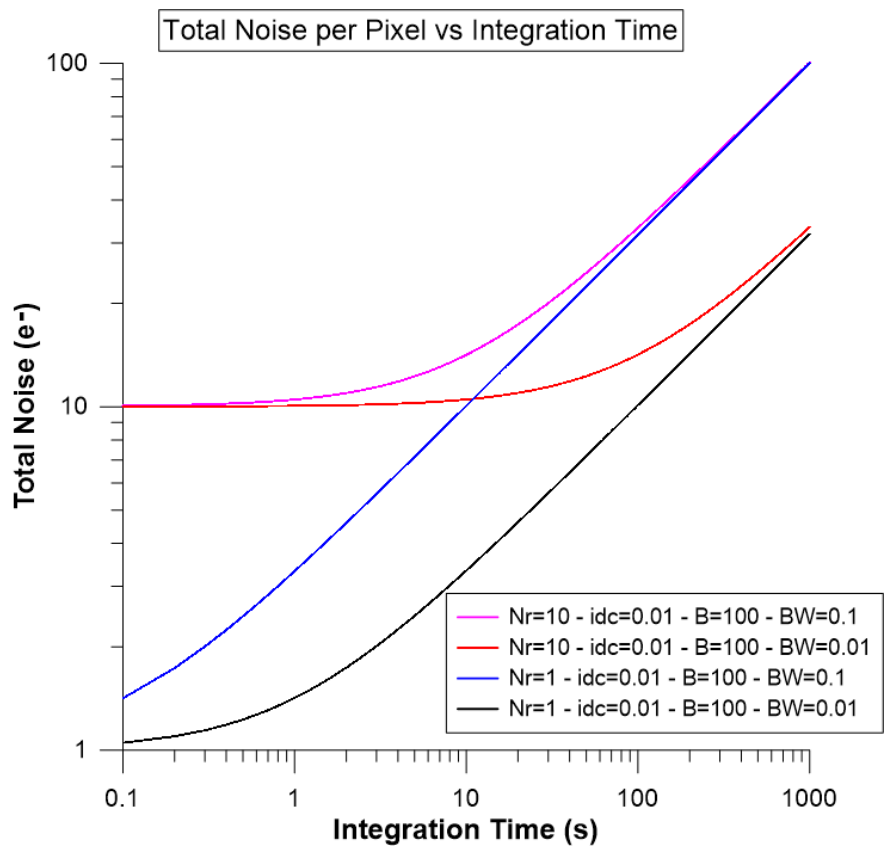

**Figure 27** – Noise per pixel (outside of signal pixels) for various detectors and bandwidths vs integration time. Background B is in ph/m²-s-micron. Bandwidth BW is in microns.

**Probability of detection of a civilization** – Assume there is **one transmitting civilization** $S_t$ in the universe and **one detecting civilization** $S_r$. We can compute the probability of detection by the receiving civilization. In order to have a detection the signal must arrive after the receiving civilization has evolved to the point of being able to detect the signal. We assume the detecting civilization is at luminosity distance L from the transmitting civilization and thus the probability of detection is the same as the fraction of sky covered by the transmitting civilization assuming a random distribution for the transmitting and receiving civilizations. However, the receiving civilization needs to be receiving during the time the transmitting civilization transmitted beam arrives. If the receiving civilization integrates for longer than the required time to the desired SNR τ then the "probability of detection" is unity IF the receiving civilization is pointed at the transmitting civilization and the transmitting civilization was pointed at the receiving civilization, modulo the time of flight. The actual probability of detection $\varepsilon_{prob-det}$ assuming the transmitting civilization is "on – ie the signal could have arrived" during the time the receiving civilization is receiving AND the SNR condition is met is then:

$$\varepsilon_{prob-det} = f_t * f_r = \Omega_t \Omega_{rec}/16\pi^2 = (t/\tau) \Omega_{rec} \Omega_{beam}/16\pi^2 = \Omega_{rec}(t/\tau) 4\lambda^2 10^{-2S}/16\pi^2$$

The probability can exceed unity in this definition which simply means the signal is detected more than once. We need to think about the evolution of both the transmitting and receiving civilizations, since neither is likely to be static. If our civilization is any indication we have been unable to receive for about a billion years after life evolved and a million years after humans evolved. Recently, we entered an exponential phase of both detection and transmission capability (even if not utilized) with doubling times of under 2 years. This represents a fundamental complexity in analyzing even our own civilization since the time scale of technological evolution is now vastly shorter than "natural time scales" such as the Sun's lifetime (Gyrs) or the time to "start" technological expansion (Myrs for "human" life). While we naturally focus on the present, it is not reasonable given the extremely small fraction this represents. If we even project 100 years into the future at our current pace we will be in a radically different place to receive and transmit. If our current doubling time persists for this 100 years and assuming a 2 year doubling time, the increase in power would be a factor of $2^{50} \sim 10^{15}$ or a civilization class change of $\Delta S \sim 7.5$. While our current construction capability (not the same folding time as photonic and electronic capability) may limit us currently, this too could change. Such an enormous civilization change would rapidly push us to ponder other limits such as the power of a star to drive a system and thus other saturation effects will no doubt evolve as our technology evolves.

**Intelligent Targeting and Filling Factors** – Based on our limited (to one) knowledge of life it makes sense, IF we were the civilization transmitting, to target individual "high value" targets such as individual stellar systems or galaxies rather than "empty space". For example in our galaxy there are approximately $10^{11}$ stars. Until we know more about the probability distribution of likely stellar candidates we could simply target individual stars and stellar systems instead of just uniformly spreading the transmission time. The covering fraction of "solar systems" cross sections is extremely small compared to the total galactic cross section. We will assume 1 AU for a solar system radius to start (we can scale from there). A simple estimate of the "solar systems" cross section is (number of stars)* (area of solar system). The covering fraction is ~ "solar systems" cross section/ (diameter galaxy)$^2$. The covering fraction then is ~ $10^{11}$ (# stars in our galaxy) x $(3\times10^{11}$ m (diam Earth orbit)$)^2$ /$(10^5$ (ly) x $10^{16}$ m/ly$)^2 \sim 10^{-8}$. Even if we expand a planetary radius to 10 AU (~ Saturn) the covering fraction is still only ~ $10^{-6}$ though we expect the typical "habitable zone" to be smaller than 10 AU. Ideally we would target individual planets (assuming this is where life exists) IF we knew where the planets were AND where to point so they transmitted signal were to intercept the planets upon arrival (we would need to understand the galactic ephemeris and gravitational lensing). As an example, a Class 4 system has a beam size of about $4.5 \times 10^{-20}$ st or about $2.8 \times 10^{20}$ beams (gain) on the sphere ($4\pi$). As we have $10^{11}$ stars and more than $10^{20}$ beams we can gain a factor of more than $10^9$ ($=2.8 \times 10^{20}$ beams/$10^{11}$ stars) by using intelligent targeting of the stars rather than the "empty space" in between. This, of course, assumed we would only target stars in our galaxy and not the distant galaxies beyond which may well be in this "empty space" between the stars in our galaxy. Similarly a Class 3 civilization has about $2.8 \times 10^{18}$ beams and we can gain a factor of more than $10^7$ ($=2.8 \times 10^{18}$ beams/$10^{11}$), while a Class 2 civilization has about $2.8 \times 10^{16}$ beams and we can gain a factor of more than $10^5$ ($=2.8 \times 10^{16}$ beams/$10^{11}$). **This makes a dramatic difference in the probability of detection as shown.** IF the transmitting civilization has knowledge of the planets around distant stars then the "intelligent targeting gain factor" is roughly correct. To understand this more we must consider the beam size at the distant system. For example, the fully synthesized beam for a Class 4 system is about $2 \times 10^{-10}$ rad. At the "edge" of our galaxy ($\sim 10^5$ ly) this corresponds to a spot size of about $10^{11}$ m. This is about 1 AU or far larger than any known planet but smaller than our solar system. A Class 3 system has a beam 10 times larger or about 10 AU at the "edge" galaxy. It is important to consider the "filing factor" of the distant solar systems IF the transmitting civilization lacks detailed knowledge about the planets and their orbits. In this case the best approach would be to "raster scan" the "stellar system" out to a "reasonable distance" away from the star in order to intercept high value (possible) planets. This might be 1-10 AU for example, depending on knowledge of the stellar class and likely "habitable zones".

**Independence of Average Deposited Energy on Planets with Target Distance** – If we assume a population of exo-planets where that the average orbital radius is $r_o$ then the average energy $E_{dep}$ deposited in time $\tau_{dep}$ within the orbital radius with the target exo-planet a distance L away from the transmitting civilization, and hence on the planet with unknown position (phase) is roughly $E_{dep} = P \tau_{dep} / \pi r_o^2$ **as long as the beam size at the beam at the distance L is smaller than the orbital diameter. Note that this statement is INDEPENDENT of the target distance as long as the beam is smaller than the orbital diameter.** If the transmitting civilization adopts the "Intelligent Targeting" strategy and places equal energy and hence spends equal time $\tau_{dep}$ per stellar system then this statement is the key to why this general technique works. The fact that we can currently (or will soon do so) achieve sub nano radian beam means even at the edge of our galaxy ($10^5$ ly) we can achieve ~ 1 AU beam diameters for a class 4 civilization operating at 1μm or at 1 Mly (roughly the distance to the nearest galaxies) we can achieve ~ 10 AU beams (class 4) and hence even at nearby extra galactic scales we have beams that are smaller than our solar system. The transmitting civilization will likely have figured this out as well and hence targeting of exo-planet solar systems becomes feasible.

**Comparing optical and radio techniques** – In comparing radio and optical/IR techniques we have to keep in mind several considerations. Some are fundamental and some are related to current technologies. In addition since we have no idea what "they" are "thinking" we cannot make any really definitive statement in this area so this is always a "conversation stopper" in SETI discussions. Nonetheless we continue.

In the terms of the flux in power and photon units at the target at distance L for class S operating at wavelength $\lambda$ we have:

$P(w)/\Omega(st) = F_E \varepsilon_c 10^{2S}/4 \lambda^2 10^{-2S} = 1400 \varepsilon_c 10^{2S}/4 \lambda^2 10^{-2S} = 350 \varepsilon_c \lambda^{-2} 10^{4S}$

$F(w/m^2) = P/L^2 \Omega = F_E \varepsilon_c 10^{2S}/(L^2 4 \lambda^2 10^{-2S}) = F_E \varepsilon_c 10^{4S}/(4L^2\lambda^2)$

$F(\gamma/s\text{-}m^2) = \xi P/L^2 \Omega = \xi F_E \varepsilon_c 10^{2S}/(L^2 4 \lambda^2 10^{-2S}) = \xi F_E \varepsilon_c 10^{4S}/(4L^2\lambda^2) = (hc)^{-1} F_E \varepsilon_c 10^{4S}/(4L^2\lambda)$

Note that $F(w/m^2) \sim 1/\lambda^2$ while $F(\gamma/s\text{-}m^2) \sim 1/\lambda$

With our current technology we can count photons in the optical and near IR but not currently in the radio. This is a technological but not fundamental limit. Even if we could count photons in the radio we would still need to build much larger (by the (ratio of the wavelengths)$^{1/2}$) systems in the radio than in the optical to achieve the same photon flux on the target **for the same power emitted**. To achieve the same power flux at the target we would need to build a larger radio array that is larger by the ratio of the wavelengths **for the same power emitted**.

As an example if we compare the optical/ IR techniques we are currently pursuing at wavelengths near 1μm with radio SETI at (say roughly) 3 GHz or 10 cm wavelength (this is meant to be approximate and hence includes studies near the "water hole") the wavelength ratio is $10^5$. This would mean that to achieve the same power flux (not photon flux) the telescope or array would need to be $10^5$ times larger. Comparing to a class 4 ($10^4$ m array) operating at 1μm wavelength this would lead to a radio array of $10^9$ m is size or about 100 times larger than the radius of the Earth. If we were interested in equivalent photon fluxes we would need to be larger by $10^{2.5}$ or have a radio array that is $3\times10^6$ m in size. **BUT – there is a critical issue.** In order to implement "Intelligent Targeting", which is critical to increasing the probability of detection, the beam size needs to be small enough to fit within the orbital diameter of the exo-planet. This would force one to build radio transmitters that are extremely large (roughly by the ratio of the wavelengths). This is why radio surveys cannot cover much of the galaxy in a survey that is power flux limited at the same level as an optical survey for the same civilization class. Nonetheless it is extremely important to cover as much of the EM spectrum as we can and hence radio surveys are critical to continue.

For example the beam size of Arecibo (300 m diameter) operating at 3 GHz or 0.1 m wavelength (near upper limit of Arecibo) would produce a beam size of about 1 mrad with is equivalent to an optical system operating at 1μm with a diameter of 3 mm. At a distance of even 1 ly the Arecibo beam would have a size of about $10^{13}$ m or about 70 AU. A "class 4 Arecibo operating at 3 GHz" with a 10 km size (the HSKA (Hundred Square Kilometer Array)) would have a beam at 1 ly distance of about 2 AU. A class 4 Arecibo at the nearest star (Alpha Centauri – 4.4 ly) would produce a beam size at the Earth of about 9 AU and produce a power into an Earth Arecibo of about 5 nW (easily detected) while the same class 4 operating at 1μm would produce a spot of about the radius of the Earth and a power of about 0.5 mW (a small laser pointer eq) into a modest 1 m optical telescope on Earth. The optical signal is vastly brighter than the brightest star in the sky and about the brightness of the full moon in a "point source". It would have magnitude about -13. It is easily seen in a cell phone camera (assuming it could detect at 1μm). Both assume 100 GW transmission.

**Simple Beacon and Search Strategies** – If another civilization adopts this "intelligent targeting" strategy and leaves the beacon on long enough we can **show that such searches have unity probability of detecting even a single comparably advanced civilization anywhere in our galaxy within a relatively short search time (few years).** This assumes that civilization is beaconing at a wavelength we can detect and that civilization left the beacon on long enough for the light to reach us now. In this blind beacon and blind search strategy the civilization does not need to know where we are nor do we need to know where they are. The civilization must understand the galactic ephemeris, in particular transverse or proper motion and understand some reasonable level of gravitational deflection and lensing in the galaxy. This same basic strategy can be extended to extragalactic distances. In figures 28, 29 and 31, 32 we show some simple ground based searches using very modest assets using 0.1 and 1 m telescopes. The key issue is that the civilization must understand the concept of "intelligent targeting" to optimize detection.

**Nearby Extragalactic Survey** – There are 127 galaxies within about 12 Mly of the Earth. Among these are a number of large galaxies including Andromeda (M31) which being the closest (large galaxy) is about 2.5 Mly away. Andromeda contains approximately one trillion stars or at least 2-4 times the number of stars as our galaxy. A class 4 civilization on Andromeda has an equivalent photometric magnitude of approximately $m_v$=17. This is easily detectable in a small (20 cm diameter) consumer telescope with a low cost camera integrating for less than 100 seconds. The dominant stellar population of Andromeda has an angular size of about 2-3 degrees. This is a convenient size that can be surveyed with either a wide field telescope or a raster scan of narrowed images. In a single 1 square degree image of the core region of Andromeda we could survey more than 100 billion stars in a single image and thus close to that many exoplanets, assuming Andromeda has a similar distribution of exoplanets as we have seen in our own galaxy with Kepler. This is clearly an extraordinarily rich target. While the average distance to these stars is about 25 times further than the distant stars in our own galaxy and thus will have a smaller flux by the square of the distance from the same civilization class, the ability to observe this large number of potential exoplanets in one image gives a unique SETI opportunity. Quantitatively it takes less than 1 ms of exposure to the class 4 civilization beam in a 1 m telescope on the Earth to achieve an SNR=10 (Fig 26). **If a class 4 civilization in Andromeda wanted to target the Milky Way and used our "intelligent targeting" scheme to maximize detection by intelligent life on planets, such as ourselves (ie target the stars in the Milky Way), then a simple Earth based 3 year survey with a 1 meter telescope would detect a single class 4 civilization anywhere in Andromeda with near unity probability.** This is also essentially what is shown in Fig 30 and 31 – right hand panels. This assumes the Andromeda civilization is transmitting long enough for us to technologically evolve to the point where we would indeed mount a search to search for "them" and that we were receiving on a wavelength they were transmitting on. This also requires that the civilization has a detailed knowledge of our galaxy's stellar motions in order to predict where the Milky Way stars and hence planets are when the signal arrives. A class 4 beam is about 0.2 nrad for $\lambda$=1µm and at 2.5 Mly (ie spot size in the Milky Way from Andromeda) has a spot size of $5 \times 10^{12}$ m or about 33 AU. This is well matched to a solar system size. At present we do not possess the technology to predict the position of stars with this precision so this remains a question as to whether more advanced civilization would have this capability.

The SNR is relatively independent on our effective spectroscopic detection resolution and the same statement (near unity probability of detection) is true for R=1 and R=1000 modulo issues such as OH emission lines which depend on the wavelength being detected. A simple search strategy uses fixed bandpass filters with a possible multichroic splitter, among other schemes.

In addition to Andromeda there are also many other nearby galaxies with similarly target rich environments though the increasing distances decrease the probability of detection for a given civilization class and a given Earth based observing asset. There are other smaller nearby galaxies that are closer than Andromeda as well.

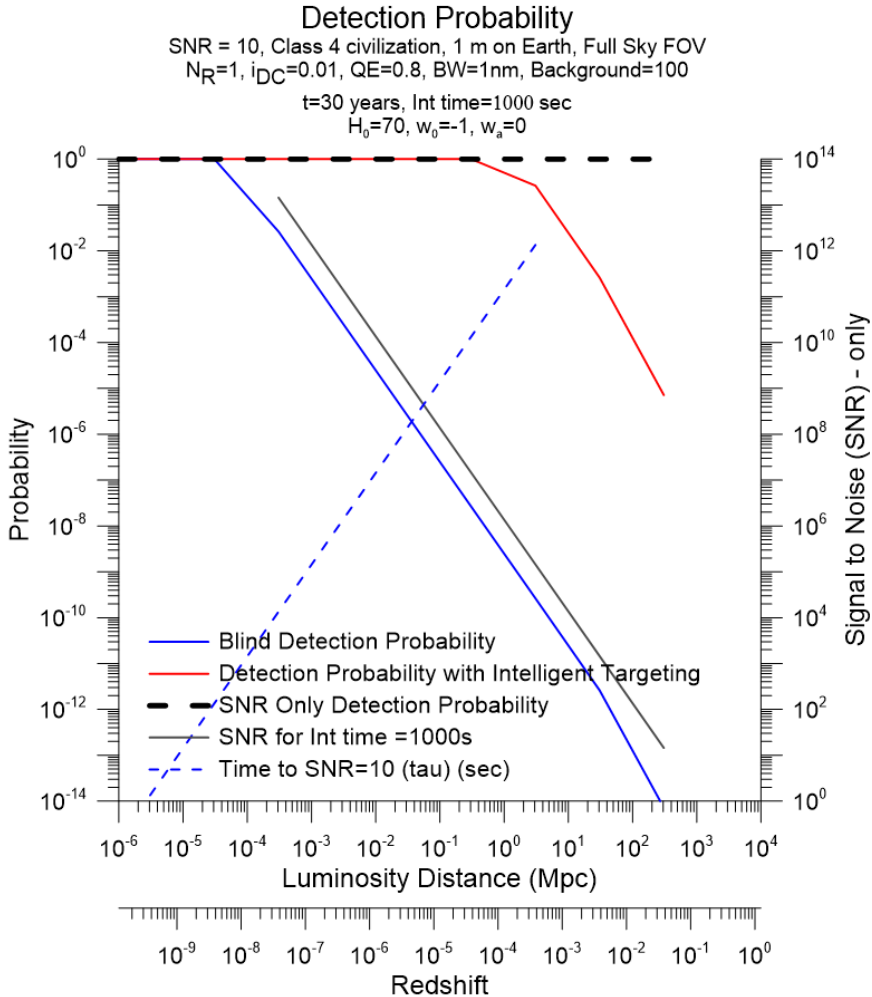

**Figure 28** – Probability of detection vs luminosity distance and redshift for a modest 1m ground based wide field survey that observes the full sky all the time for 30 years with an integration time of 1000 s per image and filter BW=1nm. The transmitting civilization is Class 4. Three types of detection probability are shown. One is based only on achieving the SNR (10 here) for a single integration time and assumes the Earth based system views the transmitter beam. The second (blue) is a blind survey of a SINGLE civilization that randomly (or uniformly) scans the sky during a 30 year Earth observing campaign. As can be seen the blind survey could easily detect the civilization if it were pointing at the Earth. The third (red) is computed assuming "intelligent targeting" is used where known stars "habitable zones" are targeted, using the "gain factor" discussed, rather than simply a uniform scan. Note that in this case the probability of detection increases dramatically. The SNR (dark grey) for integration time = 1000 s is also shown and uses the right-hand Y axis. The integration time to SNR=10 (blue dashed) is also shown – use left-hand Y axis. A benchmark or concordance model is used for the cosmological relationship between luminosity distance and redshift.

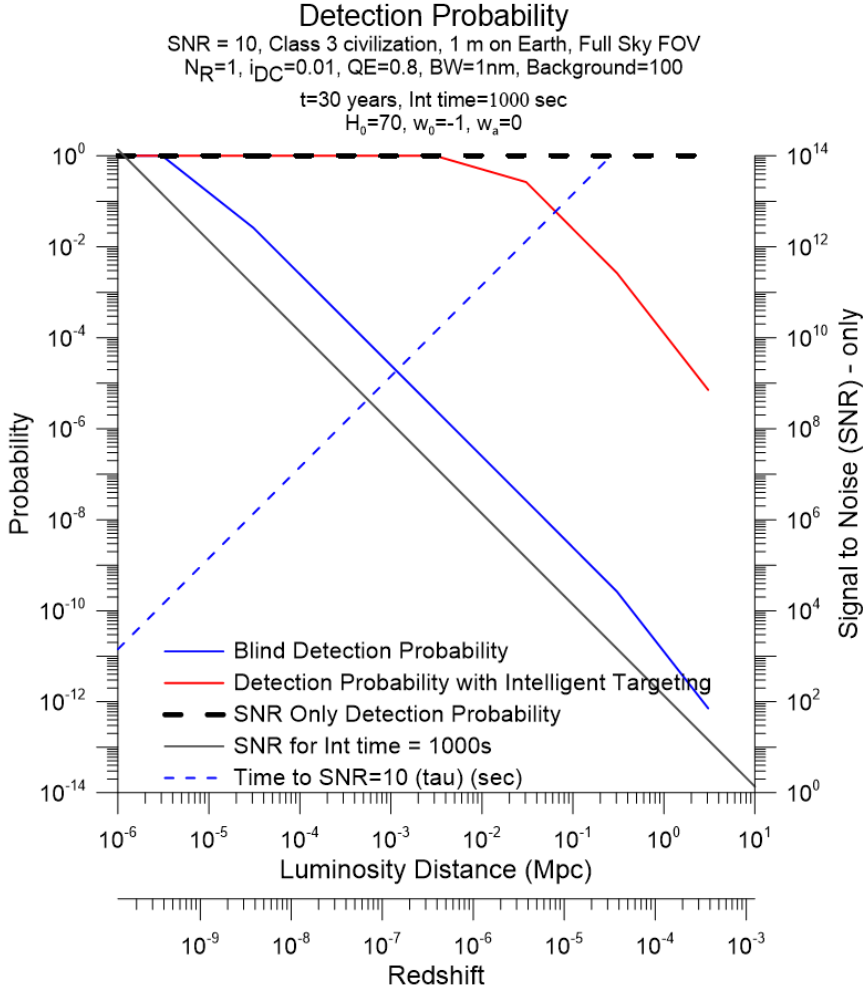

**Figure 29** – Probability of detection vs luminosity distance and redshift for a modest 1m ground based wide field survey that observes the full sky all the time for 30 years with an integration time of 1000 s per image. The transmitting civilization is Class 3 and filter BW=1nm. Three types of detection probability are shown. One is based only on achieving the SNR (10 here) for a single integration time and assumes the Earth based system views the transmitter beam. The second (blue) is a blind survey of a SINGLE civilization that randomly (or uniformly) scans the sky during a 30 year Earth observing campaign. As can be seen the blind survey could easily detect the civilization if it were pointing at the Earth. The third (red) is computed assuming "intelligent targeting" is used where known stars "habitable zones" are targeting, using the "gain factor" discussed, rather than simply a uniform scan. Note that in this case the probability of detection increases dramatically and even a class 3 civilization can be seen throughout our galaxy in an intelligently targeted transmitting survey. The SNR (dark grey) for integration time = 1000 s is also shown and uses the right-hand Y axis. The integration time to SNR=10 (blue dashed) is also shown – use left-hand Y axis.

**Multimoding into N beams** – Since we assume a phased array transmission system the beam can be split into as many beams as desired up to the number of sub elements. If they were to split into N beams the flux per split beam would be:

$$F_N (\gamma/\text{s-m}^2) = \xi P/N/(L^2 \Omega_N) = \xi P/(N^2 L^2 \Omega) = F/N^2 = \xi F_E \, \varepsilon_c \, 10^{4S}/(4L^2\lambda^2)/N^2$$

With the solid angle of each beam being $\Omega_N = N \, \Omega = 4N\lambda^2 10^{-2S}$. The flux received when split into N beams is $F_N (\gamma/\text{s-m}^2) = F/N^2$ and is reduced by $1/N^2$ since the power per beam is reduced by $1/N$ and the area covered at a given distance L is just N times larger (solid angle is N times larger). In general this will reduce the SNR by $1/N^2$. As long as the SNR is larger than the detection threshold this does not reduce the detection probability but if the SNR drops below a detection threshold then this dramatically reduces the probability. From the perspective of the transmitting civilization there may be no way to determine the receiving civilization distance nor detection capability and thus in general getting the highest SNR maximizes the detection probability. It is important to consider the SNR in the transmission strategy since this will ultimately set the number of targets for any given assumed reception capability. When we do this it is clear that the maximum number of targets, for a targeted survey, is achieved when we do not split the beam since the SNR (for a given reception capability) is proportional to the flux, which scales as $1/N^2$ and the number of targets transmitted to simultaneously is scaling as N, the product then scales as 1/N. Hence the number of detections is maximized, in general, by not splitting the beam for a phased array as opposed to a non phased array (incoherent system) where the flux (and SNR) simply adds as the number of sub elements N whereas in a phase array the flux (and SNR) adds as $N^2$.

Thus for most blind search strategies there is no advantage (and usually a large disadvantage) to having the transmitting civilization split the beam unless there is a time cadence in the receiving civilization that is relevant, though this is unknown to the transmitting civilization. For example in the "Intelligent Targeting" scheme (targeting all known stars in our galaxy for example or known galaxies for more distant targets the transmitting civilization can greatly enhance the probability of being detected. For example a Class 4 system could have $N=10^8$ simultaneous beams if each sub element (1 m in our baseline) we used for targeting. This would require about $10^3$ "transmitting exposures" to cover all the stars in our galaxy. Recall, the fully synthesized beam for a Class 4 system is about $2 \times 10^{-10}$ rad or a spot size of $10^{11}$ m (~ 1 AU) at the edge of our galaxy (~ $10^5$ ly). As this spot size is far smaller than a "solar system" that may be of interest, we can broaden the beam if needed. Depending on the transmitting civilization operational strategy (for example perhaps known "high value targets") beam splitting allows large numbers of star systems to be covered simultaneously. However, the reduction of the detected SNR being reduced by $1/N^2$ is key to factor in. Since a phased array can be into as many beams this gives a large amount of flexibility.

### 7.7 – Optical beam dwell time

An important issue to ponder is "how long would a transmitted beam be visible IF the beam was NOT tracking us"? We can make an estimate of this as follows. Assume the distance to the transmitter is L and from the point of view of the transmitter we will assume an Earth transverse speed of $v_T$. The full width beam size for a civilization class S is $\theta = 2 \lambda/d$ where $d(m)=10^S$ with $\theta = 2 \lambda 10^{-S}$ and thus the spot size "s" at the Earth is $s = L \theta = 2 L \lambda 10^{-S}$. The dwell time (Earth crossing time) $\tau = s/v_T = 2 L \lambda 10^{-S} / v_T$. Typical transverse speeds at large distances are in the $v_T =100-1000$ km/s range. This includes a typical galactic rotation speed. For reference the Earths orbital speed around the Sun is about 30 km/s and the Earths orbital speed around the galaxy is about 300 km/s. As seen in the accompanying figure the dwell time is typically long compared to our assumed putative integration time of 1000 seconds except for short distances and large civilization class. However in the latter cases the SNR would be extremely large even at spot dwell times much shorter than the 1000 sec integration time. For simplicity we assume a Euclidean geometry.

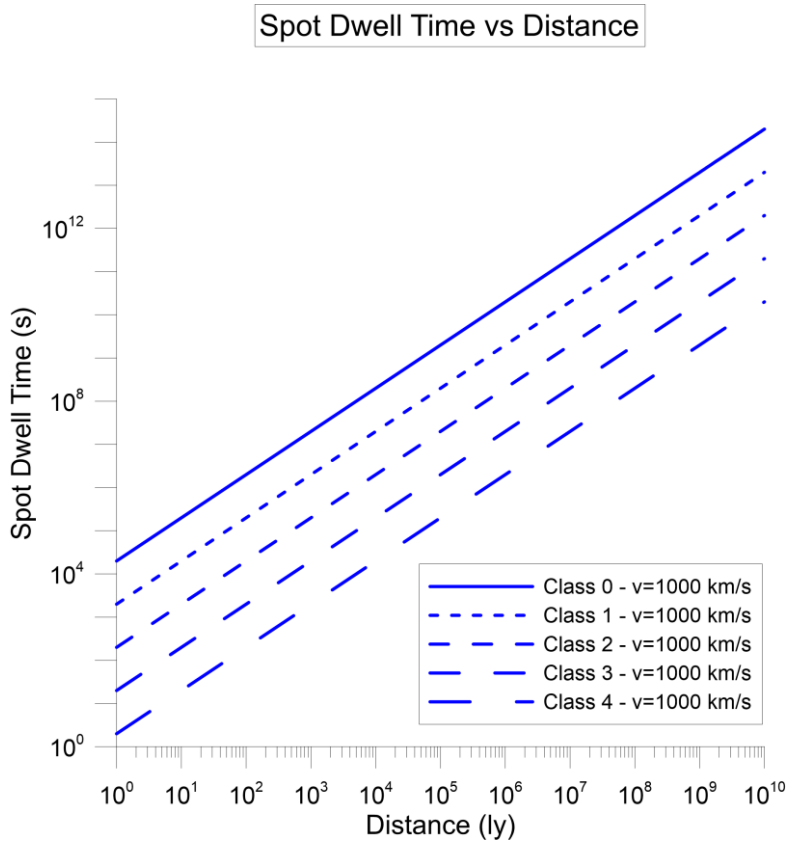

**Figure 30 – S**pot dwell time vs distance and class with an assumed comoving (relative to radial) transverse speed of 1000 km/s.

### 7.8 – The idea of "Naturalness"

One could argue on the basis of "natural wavelength windows" that one approach is "better" or more "likely" than the other. But there is no real "logic here as we have no idea what is logical to another civilization. Anyone who has observed SETI programs knows that we search with whatever our latest technology available is. As mentioned, our technological phase has only been an extremely small fraction of humanities existence, let alone life on Earth. A "reasonable" question is to ask what happens if we allow technology to mature to some modest fraction of human existence (say 50%) and then we readily see that instead of considering the last 100 years of feasible SETI ideas we might consider 1 million years of technological advancement. While we can project a roadmap into the next decade or so we certainly have extremely little predictive power into hundreds, let alone millions of years. We have to be honest and fall back to "what can we do now". What is new now is that we can now search for another similarly advanced civilization across the entire universe. This IS new to us. What is an assumption , of course, is that electromagnetic communications has any relevance on times scales that are millions of years and in particular that electromagnetic communications (which includes beacons) should have anything to do with wavelengths near human vision. We could simply "throw up our arms and give up" but this is not our nature. We proceed to explore within the limits of reasonable resource use.

## 7.9 – Communications between civilizations

The idea that any form of electromagnetic signal would be used as a form of communications is one that we are used to from our everyday lives. A major issue occurs when we extend this to long range communications where long range is measured in units of the distance between stars or galaxies. Here the time of flight (years to millions or billions of years) becomes a major point of discussion. We are used to communications being "full duplex" namely that "send and receive" or "speak and listen" happen with a delay that is very short compared to our lifetime. Even in our solar system the communication are "half duplex" in that we transmit and then must wait a significant period of time to receive a response. The idea that civilizations that are widely spaced would communicate in "real time" with each other with any form of electromagnetic signal thus seems highly illogical. As we do not have any faster way of communications (no Tachyons yet) we have a philosophical and scientific quandary as to why distant civilization would in fact use any form of "light speed" communications system except as a beacon or as a "one way" streaming of information, much like television - ie non interactive.

**Beaming vs Communications** - A more logical scenario seems to be one where civilizations search for other civilization by "beaming" out their existence and (possibly) waiting for a response over long periods of time. In essence that is what the entire SETI effort has been focused on, except we generally simply listen. Thus the idea that we will "listen in" on the communication between civilizations seems unlikely whereas the idea of civilization that pro actively broadcast their existence, such as a firefly does, seems more logical. An alternative (logical) scenario is that we will detect the beam from a civilization that uses power beaming for utilitarian purposes such as propulsion. This would require a chance detection of either an errant beam or "spillover". However in all of this "logic" is very much an anthropomorphic construct.

## 7.10 – Signals from other application of directed energy

There are a number of reasons a civilization would use directed energy systems of the type discussed here. If other civilizations have an environment like we do they might use DE system for applications such as propulsion [22], planetary defense against "debris" such as asteroids and comets [17,18,20], illumination or scanning systems to survey their local environment [19], power beaming across large distances among many others. Surveys that are sensitive to these "utilitarian" applications are a natural byproduct of the "spill over" of these uses, though a systematic beacon would be much easier to detect.

## 8. ACTIVE VS PASSIVE

In general SETI (with a few and controversial exceptions) has been carried out in a completely passive mode – ie we listen and do not speak. Perhaps we learned this as children or perhaps it is born out of fear from science fiction stories and movies. In general we have both a curiosity and a fear of the unknown. This is a natural survival instinct [23,24]. There is also a completely rational part to listening vs speaking – namely the finite speed of light. When we speak (transmit) it will take a minimum of 4 years to reach the nearest stellar system (Alpha or Proxima Centauri), 1000 years to reach the Kepler planets, more than 2 million years to reach the nearest large galaxy (Andromeda) and close to 100 million years to reach the nearest galaxy clusters. With the exception of the nearest stars, these time scales are far beyond a human lifetime and perhaps more importantly they greatly exceed the time scale for "radical technology evolution". Another issue is that all stars and galaxies have a proper (transverse) velocity relative to our line of sight. This is often of order $\beta \sim 10^{-3}$. This means that if we observe a distant star or galaxy and want to transmit to it then its proper motion will have moved it from our initially targeting of it. It will have moved by an angle of approx $\beta$ (in radians). This is an enormous angle relative to the beam size for even a modest system where the (full) beam size is $\theta = 2 \lambda 10^{-S}$. Even for an S=1 civilization (less than us) and $\lambda = 1\mu$ we have $\theta = 2\mu$rad which is much smaller than a typical proper motion $\beta$. In order to hit the target we would have to have detailed knowledge of the dynamics and integrated gravitational field as well as gravitational lensing along the way. This is not a trivial task and one where civilizations may resort to beam broadening or multi beam transmission to increase detection probability. Depending on the detection temporal strategy these transmission strategies may not increase the detection probability. It is a complex mix of SNR for a given civilization transmission class and civilization reception class. **Our hope in SETI is that other advanced civilizations, if they exist, are not as scared as we are to transmit, otherwise the silence in the universe will be deafening.**

## 9. CONCLUSIONS

We have now reached the point in human technological evolution to project our own presence across the entire universe. The question is "are there other civilizations for which this is also true"? If so are they now signaling us? We have shown that even our current technology is capable of being detected across virtually the entire horizon if we chose to do so and that we are on an extraordinarily rapid ascent phase in this technology. We have shown that even modest directed energy systems can be "seen" as the brightest objects in the universe within a narrow laser linewidth. We have outlined logical search strategies that search for signatures of an exceeding large number of candidates on cosmological scales, including searches at high redshift, that can help us search for the answer to the question of "are we alone". This can be done with very modest resource allocations.

## ACKNOWLEDGEMENTS

We gratefully acknowledge funding from the NASA California Space Grant NASA NNX10AT93H in support of this research and NASA NIAC 2015 NNX15AL91.

# Appendix
Additional cases

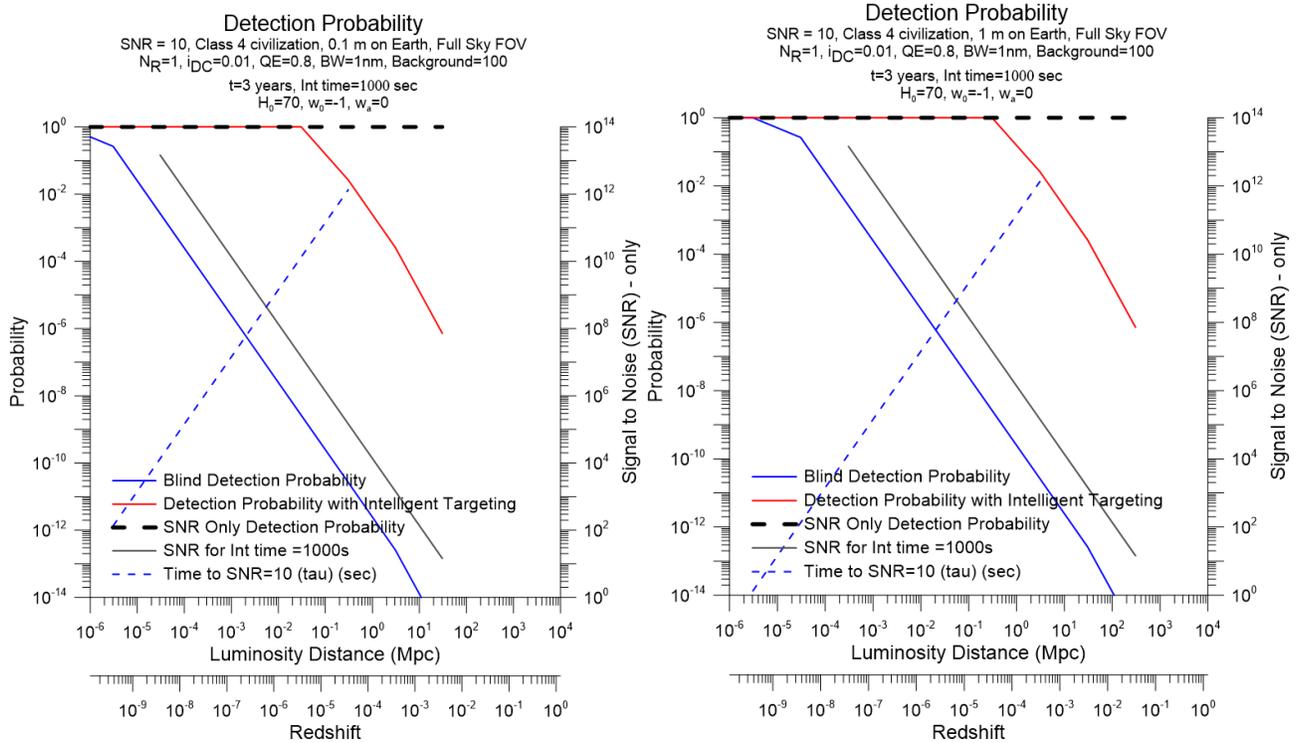

**Figure 31** – **Left:** Probability of detection vs luminosity distance and redshift for a small 0.1m ground based wide field survey that observes the full sky all the time with an integration time of 1000 s per image and filter BW=1nm for 3 years. The transmitting civilization is Class 4. The SNR (dark grey) for integration time = 1000 s is also shown and uses the right-hand Y axis. The integration time to SNR=10 (blue dashed) is also shown – use left-hand Y axis. With intelligent targeting of stellar systems even a 0.1 m full sky survey will detect a class 4 civilization anywhere in the galaxy assuming that the civilization randomly beacons and uses the intelligent targeting strategy as discussed in the text. **Right:** Same for 1 m. Note that the Intelligent Targeting probability of detection is virtually identical whether the integration time is 1 or 1000 seconds.

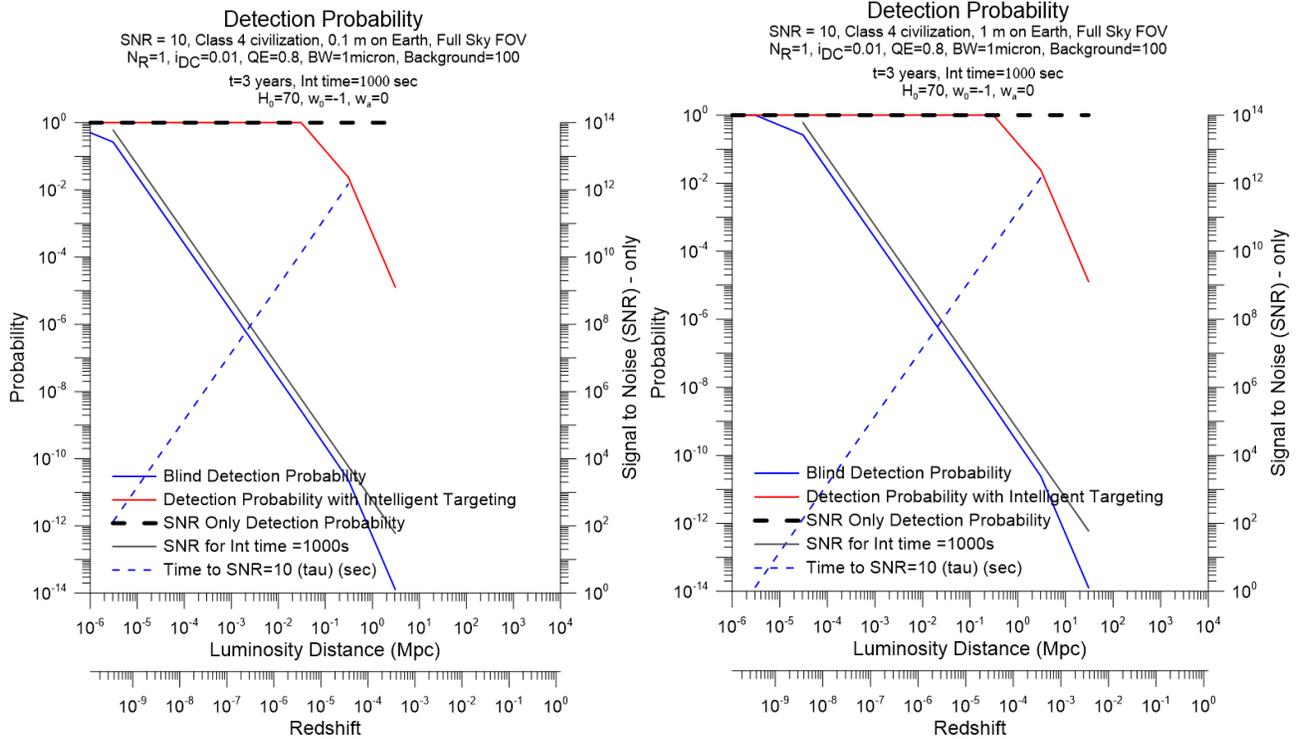

**Figure 32** - **Left:** Probability of detection vs luminosity distance and redshift for a small 0.1m ground based wide field survey that observes the full sky all the time with an integration time of 1000 s per image and wide filter BW=1micron for 3 years. Note that in the near IR OH lines would need to be included with such a wide bandwidth and this will depend on the precise spectral coverage. This is NOT included here. The primary point is that modulo OH lines the detection probability is relatively insensitive to detection bandwidth BUT spectral specificity is needed for systematic reasons in general. The transmitting civilization is Class 4. The SNR (dark grey) for integration time = 1000 s is also shown and uses the right-hand Y axis. The integration time to SNR=10 (blue dashed) is also shown – use left-hand Y axis. With intelligent targeting of stellar systems even a 0.1 m full sky survey will detect a class 4 civilization anywhere in the galaxy even with very wide filter bandwidth assuming that the civilization randomly beacons and uses the intelligent targeting strategy as discussed in the text. **Right:** Same for 1 m. Note that the Intelligent Targeting probability of detection is virtually identical whether the integration time is 1 or 1000 seconds.